\documentclass[showpacs,twocolumn,aps,preprintnumbers,letterpaper]{revtex4}
\usepackage{amsmath,amssymb}
\usepackage{epsfig}
\usepackage{graphicx}
\usepackage{amsmath}
\usepackage{slashed}
\usepackage{amsfonts}
\usepackage{epstopdf}

\newcommand{\be}{\begin{equation}}
\newcommand{\ee}{\end{equation}}
\newcommand{\bear}{\begin{eqnarray}}
\newcommand{\eear}{\end{eqnarray}}
\newcommand{\ba}{\begin{array}}
\newcommand{\ea}{\end{array}}

\def\be{\begin{eqnarray}}
\def\ee{\end{eqnarray}}
\def\bea{\be}
\def\eea{\ee}

\def\roughly#1{\mathrel{\raise.3ex\hbox{$#1$\kern-.75em%
\lower1ex\hbox{$\sim$}}}}

\begin{document}

\title{Confining Dyon-Anti-Dyon  Coulomb Liquid Model I}

\author{Yizhuang Liu}
\email{yizhuang.liu@stonybrook.edu}
\affiliation{Department of Physics and Astronomy, Stony Brook University, Stony Brook, New York 11794-3800, USA}

\author{Edward Shuryak}
\email{edward.shuryak@stonybrook.edu}
\affiliation{Department of Physics and Astronomy, Stony Brook University, Stony Brook, New York 11794-3800, USA}

\author{Ismail Zahed}
\email{ismail.zahed@stonybrook.edu}
\affiliation{Department of Physics and Astronomy, Stony Brook University, Stony Brook, New York 11794-3800, USA}


\date{\today}
\begin{abstract}
We revisit the dyon-anti-dyon liquid model for the Yang-Mills confining vacuum discussed by Diakonov and Petrov, by
retaining the effects of the classical interactions mediated by the streamline between the dyons and anti-dyons.  In the SU(2) case the
model describes a 4-component strongly interacting Coulomb liquid in the center symmetric  phase.
We show that in the linearized screening approximation the streamline interactions  yield
 Debye-Huckel type corrections to the bulk parameters such as the pressure and densities, but do not alter significantly the
large distance behavior of the correlation functions in leading order. The static scalar and charged structure
factors are consistent with a plasma of a dyon-anti-dyon liquid with a  Coulomb parameter $\Gamma_{D\bar D}\approx 1$
in the dyon-anti-dyon channel.
Heavy quarks are still linearly confined and the large spatial Wilson loops still exhibit area laws in leading order. 
The t$^\prime$ Hooft loop is shown to be 1 modulo Coulomb corrections.
\end{abstract}




\maketitle

\setcounter{footnote}{0}


\section{Introduction}

At asymptotically high temperature $T$, QCD-like theories are in a weakly
coupled state known as the Quark-Gluon Plasma (QGP). In it semi-classical solitons -- instantons and their 
constituents, monopoles etc -- have large action $S=O( 1/\alpha_s)\gg 1$. Their semi-classical
treatment is parametrically reliable, but their density is exponentially suppressed by $e^{-S}$. As a
result their effects are small.

However, as the temperature decreases the semi-classical action $S$ decreases. Since the soliton
density grows as a power of $1/T$ their contribution to the QCD partition increases. 
 At a critical density fixed by $T_c$,  confinement sets in, and the near-zero expectation value
of the Polyakov line $\left<L\right>\approx 0$ switches off the quark component of the QGP, as well as the
(non-diagonal) gluons. Below the critical temperature $T_c$, the solitons dominate the field ensemble.

The major questions at the transition point are: (i) Are these objects  still made of strong enough fields,  
allowing for a semi-classical analysis; (ii) Are  their interactions weak enough to preserve their 
individual identity;  (iii) Are the semi-classical interactions in the thermal ensemble amenable 
to known methods of many-body theory. As we will argue below, two first questions
will be answered in the affirmative, and the third also, provided the ensemble is dense enough.

The instanton liquid model developed in the 1980$^\prime$s is an example of such a semi-classical treatment.
In vacuum at $T=0$, the action per typical SU(3) instanton was found to be large with $S\sim 12$,
 and the inter-instanton and anti-instanton interactions tractable. The non-perturbative 
 vacuum topological fluctuations are related to the explicit violation of the axial U(1), and
the formation of fermionic zero modes. The collectivization of the fermionic zero modes leads to 
the spontaneous breaking of flavor chiral symmetry~\cite{ALL} (and references therein). More recently, 
instanton-induced effects were found to be important
for hadronic spin physics~\cite{SPIN}.

However, around the critical temperature $T\sim T_c$, 
instantons should know about the non-vanishing of the Polyakov
line expectation value,  also referred to as a non-trivial holonomy. 
Instantons with non-trivial holonomies were found in~\cite{KVLL}.
The key discovery was that large holonomies split 
instantons  into $N_c$ constituents, the selfdual instanton-dyons.
Since these objects have nonzero Euclidean electric and magnetic charges and source
Abelian (diagonal) massless gluons, the corresponding ensemble is 
an ``instanton dyon plasma"

We remark that the  electric dyon field is real in Euclidean space-time but
imaginary in Minkowski space. The instanton-dyons are also referred to as
instanton-monopoles or instanton-quarks. However, the notion of a non-zero holonomy and all the instanton-related
contructions do not exist outside of the Euclidean finite-$T$ formulation. On the lattice, both the electric and magnetic charges
of the instanton-dyons are observable by standard Gaussian surface integrals.

 Diakonov and Petrov~\cite{DP} 
 emphasized that, unlike the (topologically protected) instantons, the dyons interact directly with
 the holonomy field. They
 further suggested that since such dyon (anti-dyon)  fields become significant
at low temperature, they  may be at the origin of a vanishing of the mean Polyakov line, or confinement.
This mechanism is similar to the  Berezinsky-Kosterlitz-Thouless-like transition
of instantons into fractional ``instanton quarks" suggested earlier by Zhitnitsky and others~\cite{ZHITNITSKY},  inspired by
the fractionalization of the topological charge in 2-dimensional CPN models~\cite{FATEEV}, although it
is substantially different in  details.
It is also different from the random dyon-anti-dyon ensemble suggested earlier by Simonov
and others~\cite{SIMONOV}. It is not yet clear  how this Euclidean mechanism relates to the
the quantum condensation of magnetic monopoles suggested initially by t$^\prime$ Hooft~\cite{THOOFT}
and Mandelstam~\cite{MAND}, and subsequently supported in the supersymmetric model discussed by 
Seiberg and Witten~\cite{SEIBERG}. In many ways, it is similar to the 3-dimensional monopole plasma
discussed by Polyakov~\cite{POLYAKOV}.

 Unsal and Yaffe~\cite{UNSAL1} , using a double-trace
deformation of Yang-Mills at large $N$  on $S^1\times R^3$, argued  that it  prevents the spontaneous 
breaking of center symmetry. A similar trace deformation was used in the context of two-dimensional (confining) QED with unequal
charges on $S^1\times R$~\cite{HOLGER} to analyze the nature of center symmetry and its spontaneous breaking. 
This construction was extended to QCD with adjoint fermions by Unsal~\cite{UNSALALL},
and  by Unsal and others~\cite{UNSAL} to a class of  deformed supersymmetric theories with soft supersymmetry breaking. 
 While the setting includes a compactification on a small circle, with  weak coupling and
 an exponentially  $small$ density of dyons, the minimum at the confining holonomy
  value is induced by the repulsive interaction in the dyon-anti-dyon pairs (called  
 $bions$ by the authors). The supersymmetry is needed to calculate the contribution of the   dyon-anti-dyon pairs ,
 and, even more importantly, for the 
cancellation of the perturbative Gross-Pisarski-Yaffe-Weiss (GPYW)  holonomy potential \cite{WEISS}.

 Shuryak and Sulejmanpasic \cite{TIN}
 have  argued that  induced by the ``repulsive cores" in dyon-antidyon channel  also generates confinement,
 explaining it in a simple model.  The first numerical study of the 
 classical interaction of  the dyons with anti-dyons has been recently carried in~\cite{LARSEN-SHURYAK}.
 The streamline configurations were found by a gradient flow method, and their action assessed. 
 This classical interaction will be included -- for the first time -- in our paper.

Another major non-pertubarive phenomenon in QCD-like theories is spontaneous chiral symmetry breaking.
 Shuryak and Sulejmanpasic~\cite{SHURYAK} have  analyzed a number of  phenomena induced by
 the fermionic zero modes of the instanton-dyons such as the formation of clusters  
 (molecules or bions) at high temperature and their collectivization, generating spontaneous breaking of chiral symmetry 
 at low temperature. Faccioli and Shuryak~\cite{FACCIOLI} have started numerical simulations of the
dyon ensemble with light fermions to understand the nature of the fermionic collectivization. We will
provide an analytical analysis of these effects in the second paper of this series.

Before we get into the details of the various approximations to our analysis,
let us try to provide some qualitative answers 
to the three generic questions formulated above: (i) At $T\sim T_c$,  we will consider the action per dyon 
(anti-dyon) to be still large or $S\sim 4$ whatever $N_c$;
 (ii) The dyon  interactions will be of the order of 
$\Delta S_{\rm int}\sim 1 \ll S$. The quantum (one-loop) interactions are several times smaller and naively can be considered small.
However they are quite non-trivial and the repulsion they provide would be our key finding.
(iii) In general, the dyon plasma is strongly coupled and it is hard to treat it analytically.
However we will argue below that in some window of temperatures (below $T_c$)   
one can still use the Debye-Huckel plasma theory.

A major contribution to the understanding of the one-loop dyon interaction has been made by Diakonov and others~\cite{DP,DPX}.
They have found that at $T>T_c$ their interaction with the surrounding  QGP leads to a linear (confining) potential
between the dyons, proportional to the perturbative Debye mass. Since in this work we will only
consider the opposite case $T<T_c$, this will not be included in what follows. Key to the one-loop effect is 
 the explicit quantum weight of the KvBLL instantons in terms of the collective coordinates of the
constitutive dyons at all separations. The self-dual sector is characterized by a moduli space with a hyper-Kahler metric.
Its volume element is given by the determinant of Coulomb-like matrix. We will refer to it as Diakonov determinant.


In his first attempts to treat the dyonic plasma, Diakonov kept only the one loop determinant,
the volume of the moduli space, ignoring the
QGP screening effects and -- as we will discuss in detail -- the even larger classical 
dyon-anti-dyon interaction. Furthermore, 
he assumed that the attractive and repulsive terms 
induced by the determinant cancel out on average. We disagree on this conclusion as we detail below.
Indeed, Bruckmann and others in~\cite{LATTICE}
tried to generate configurations of randomly placed dyons
using the determinantal measure, and observed that for the physically 
relevant dyonic densities, the determinantal measure develops negative eigenvalues. This 
makes no sense if the measure is to account for the volume of the dyonic moduli space.
We will show that this issue may become  resolved in a strongly correlated ensemble.

It is well known that the separate treatment of self dual and anti-selfdual sectors is only justified
in the context of supersymmetry where self-duality is dual to holomorphy.
In QCD-like theories, the interaction between   self dual and anti-selfdual sectors is strong
 and not factorizable. It is described semi-classically by a ``streamline" with a
 classical inter-particle potential of order $1/ \alpha_s$,  which is larger than the 
 1-loop quantum induced potential of order $\alpha_s^0$. Furthermore, configurations with too strongly
 overlapping objects with small action, are not subject to the semiclassical treatment. To account for that
 one usually relies on the use of a ``repulsive core" as in the instanton liquid model for instance.

As we will discuss in detail, classical  dyon-antidyon interaction
 \cite{LARSEN-SHURYAK} 
 is about an order of magnitude stronger than the one-loop Coulomb effects. It
  generically leads to the dyon plasma in the strongly coupled regime, with $e^{-V_{D\bar D}}\gg 1$. 
We will however focus on the very dense regime of such plasma, in which
screening is strong enough that statistical mechanics of the ensemble can be treated by a variant of the  
 Debye-Huckel mean field plasma theory.
 In such case the screening length is short enough 
to fence the system from strong coupling correlations and 
molecular-type instabilities  induced by the streamline.
The more dilute systems such as those appearing at $T>T_c$, 
will not be discussed in this work, as they need more powerful many-body methods, such as e.g. 
strongly coupled
Coulomb plasmas  many-body physics re-summations~\cite{DH,CHO} (and references therein).
As we will show, in this case the free energy has a minimum at the ``confining" holonomy value 
 $v=\pi T$.

In this paper we will detail the strongly coupled nature of the dyonic plasma. Our original results
consist of (i) introducing the strong  correlations between dyons and anti-dyons as described
by the streamline~\cite{LARSEN-SHURYAK}; (ii) showing that the determinantal interactions induced
by the moduli space for dyons or anti-dyons are mostly repulsive causing the moduli volume to vanish
for randomly distributed dyons; (iii)  showing that suitably organized dyons to account for screening
correlations yield finite moduli volumes; (iv)  deriving an explicit 3 dimensional effective action that
account exactly for the screening of dyons and anti-dyons on the moduli space with  strong
inter-dyon-anti-dyon streamline  interactions; (v) showing explicitly that the strongly coupled dyonic 
plasma is center symmetric and thus confining; (vi) deriving the Debye-Huckel corrections induced
by the dyons and anti-dyons to the leading Pressure for the dyonic plasma and using it
to asses the critical temperature for the SU(2) plasma; (vi) providing the explicit
results for the gluon topological susceptibility and compressibility near the critical temperature in the
center symmetric phase;  (vii) deriving the scalar and charged structure factors of the dyonic plasma
showing explicit screening of both electric and magnetic charges at large distances
with explicit predictions for the electric and magnetic masses; (viii) showing that 
the strongly coupled dyonic plasma supports both electric and magnetic confinement.

This paper is organized as follows: 
In section 2 we  review the key elements of the dyon and anti-dyon measure derived
in~\cite{DP,DPX} using the KvBLL instanton. The dyon-anti-dyon  measure is then composed of the
product of two measures with streamline interactions between the dyons and anti-dyons.
We briefly detail the exact re-writing of the 3-dimensional grand-partition function in terms of
a 3-dimensional effective theory in the SU(2) case. We also show that the ground state of this
effective theory is center symmetric. In sections 3-6  we show that in the linearized screening
approximation the dyon-anti-dyon liquid  still screens both electric and magnetic charges, generates a
linearly rising potential between heavy charges and  confines the large spatial Wilson loops.
The t$^\prime$ Hooft loop in the dyon-anti-dyon ensemble is shown to be 1 modulo 
${\cal O}(\alpha_s)$ self-energy corrections which are perimeter-like  in section 7.
Our conclusions are in section 8.

\section{Interacting Dyon-Anti-Dyon Ensemble}
\subsection{The setting}

The first step is the introduction of the nonzero expectation value of the 4-th component
of the gauge field, which is gauge invariant since at finite temperature it enters the holonomy
integral over the time period, known also as the Polyakov line. Working in a gauge in which
$\left<A_4\right>$ belongs to the diagonal and traceless sub-algebra of $N_c-1$ elements, one 
observes the standard Higgsing via the adjoint field. All gluons except the diagonal ones 
become massive. 
We will work with the simplest case of two color gauge theory $N_c=2$, in which
there is only one diagonal matrix and the
VEV of the gauge field (holonomy) is normalized as follows 

 \be 
 \left<A^3_4\right> =v{\tau^3\over 2}= 2\pi T \nu {\tau^3\over 2} 
 \ee
 where $ {\tau^3/2}$ is the only diagonal color generator of  SU(2).
At high $T$ it is trivial with $\nu\rightarrow 0$, and at 
low $T<T_c$ it takes the confining value $\nu=1/2$.
With this definition, the only dimensional quantity in the classical approximation is the temperature $T$,
while the quantum effects add to the running coupling and its $\Lambda$ parameter. Since we are 
working near and below $T_c$, we will follow the lattice practice and we use the latter as our main unit.

In the semi-classical approximation, the Yang-Mills partition function is assumed to be dominated by an interacting ensemble of
dyons (anti-dyons)~\cite{DP,DPX}. For large separations or a very dilute ensemble, the semi-classical interactions are mostly Coulombic,
and are encoded in the collective or moduli space of the ensemble. For multi-dyons  a plausible moduli space was argued starting
from the KvBLL caloron~\cite{KVLL} that has a number of pertinent symmetries, among which permutation symmetry, overall charge neutrality,
and clustering to KvBLL at high temperature. Since the underlying calorons are self-dual, the induced metric on the moduli space was shown to be
hyper-Kahler.

 The
SU(2) KvBLL instanton (anti-instanton) is composed of a pair of dyons labeled by $L,M$ (anti-dyons by $\overline L,\overline M$)
in the notations of~\cite{DP}. Specifically $M$ carries $(+,+)$ and $L$ carries $(-,-)$ for (electric-magnetic) charges, with
fractional topological charges $v_m=\nu$ and $v_l=1-\nu$ respectively. Their corresponding actions are $S_L=2\pi v_m/\alpha_s$ and  $S_M=2\pi v_l/\alpha_s$.



The statistical measure for a correlated ensemble of dyons and anti-dyons is

\bea
d\mu_{D\overline D}[K]\equiv &&\,e^{-V_{D\overline D}(x-y)}\\
&&\times\prod_{m=1}^N\prod_{i=1}^{K_m}\,\frac {f\,d^3{x_{mi}}}{K_m !}\,{\rm det}(G_{mi}[x])\nonumber
\\&&\times \prod_{n=1}^{ N}\prod_{j=1}^{{\overline K}_n}\,\frac {f\,d^3{y_{nj}}}{{\overline K}_n !}\,{\rm det}(G_{nj}[y])\,
\nonumber
\label{MEASUREDD}
\eea
The streamline interactions induced by the potential $V_{D\bar D}$  correlate
the two otherwise statistically independent dyon and anti-dyon sectors.
(Note that by the potential we mean the extra action and not the energy,  thus no extra $1/T$).
Asymptotically, 

\be
V_{D\overline D}(x-y)\rightarrow \sum_{mn,ij}\,\frac {C_D/2}{\alpha_s\,T} \,\frac{Q_{mi}{\overline Q}_{nj}}{|x_{mi}-y_{nj}|}
\label{VDD}
\ee
 is  a Coulomb-like classical interaction between dyons and anti-dyons.
 Here $x_{mi}$ and $y_{nj}$ are the 3-dimensional coordinate of the i-dyon of  m-kind
and j-anti-dyon of n-kind. 
At shorter separations the streamline stops at certain distance $a_{D\bar D}$,
we will refer to it as a ``core size". While the interaction is more complex than just electric Coulomb, it is proportional to
the electric charges $Q,\overline{Q}$. In general those are the (Cartan) roots of $SU(N_c)$ supplemented by
the affine root. They satisfy
\be
Q_{mi}{\overline Q}_{nj}\equiv -\left(2\delta_{mn}-\delta_{m,n+1}-\delta_{m,n-1}\right)
\label{QQ}
\ee
The dimensionality of $G[x]$ is
$(K_1+...+K_N)^2$ and similarly for $G[y]$. Their explicit form can be found in~\cite{DP,DPX}.  
In the SU(2) case there is only one electric charge.

The semiclassical 3-density of all dyon species $n_D\equiv n_L+n_M+n_{\bar L}+n_{\bar M}$ is

\be 
n_D ={dN \over d^3x} = \frac {C T^3 \,e^{-\frac{\pi}{\alpha_s}}}{\alpha_s^2} 
\label{NDDD}
\ee
where $C$ is a constant to be determined below (see (\ref{NDX})). (\ref{NDDD}) 
can be re-written using the asymptotic freedom formula for SU(2) pure gauge theory
 with $2\pi/\alpha_s(T)=(22/3)\, {\rm ln}\,(T/\Lambda)$
in terms of the
  scale parameter $\Lambda$.  The 
dimensionless  density 

\be {n_D \over T^3} \sim \left({\Lambda \over T}\right)^{11/3}  
\ee  
is small at high $T$ but increases  as $T$ decreases. 
With the exception of section \ref{sec_deconf}, where we will estimate the critical deconfinement temperature 
by including perturbative ${\cal O}(\alpha_s^0))$ effects in the dimensionless pressure,  
we will always assume the temperature to be small enough, so that the dyons effect are the dominant ones.
The dyon fugacity $f$ is

\be
f\approx \frac{n_D}{8\pi}
\label{FUGA}
\ee
to order ${\cal O}(n_D^{3/2})$ in the dyon density  (see below).
The absolute value of the parameter $\Lambda$ appearing in the semiclassical formulae 
can be related to standard parameters like $\Lambda_{\overline{MS}}$, but this has 
no practical value since the accuracy with which they are known is too low to give an accurate 
value of the dyonic density. In practice it is obtained from  
the fit to the lattice instanton 
data performed  in~\cite{SHURYAK} in the range $0.5<T/T_c<3$. The caloron action -- the  sum of
$S_L$ and $S_M$  -- is then writen as
\be
S_{L+M} (T)\equiv \frac{2\pi }{\alpha_s(T)}\equiv \frac{22}3{\rm ln}\left(\frac {T}{0.36\,T_c}\right)
\label{RUN}
\ee
We will use this fit as a basis for our running coupling. In particular, 
the action of the SU(2)  caloron at $T_c$ $S_{\rm L+M} (T_c)\approx 7.47$  translates to the value of the coupling
$\alpha_s(T_c)=0.84$. Since  in this paper we only work  in the confining regime of the
holonomy with all dyon actions identical, the action per dyon is about 3.75.


The repulsive linear interaction between unlike dyons (anti-dyons) found in \cite{DPX}
acts as a linearly confining force in the center asymmetric
phase,  favoring the molecular or KvBLL configuration at $T>T_c$.
This interaction stems from QGP thermal quanta scattering on the dyons.
However, 
 we will be interested in this paper in the center symmetric phase at $T>T_c$,
in which there is no QGP, 
 we do not include this interaction.

Since the classical $V_{D\bar D}\sim 1/\alpha_s$ it dominates the quantum determinants, which include Coulomb
interaction of order
$\alpha_s^0$. On this point we differ from the argument presented in~\cite{DP} regarding the re-organization of (\ref{VDD}) 
 in an extended quantum determinant.   At large relative separations
between all  particles the measure (\ref{MEASUREDD})  is exact. It is also exact when each bunch of dyons or anti-dyons coalesce into a KvBLL instanton or anti-instanton at all separations. 

The above notwithstanding, the grand-partition function associated with the measure (\ref{MEASUREDD}) 

\be
{\cal Z}_{D\overline D}[T]\equiv \sum_{[K]} \,\int d\mu_{D\overline D}[K]
\label{ZDD}
\ee
describes a highly correlated ensemble of dyon-anti-dyons which is no longer integrable in the presence of the streamline. The case $V_{D\overline D}=0$ amounts
to ${\cal Z}_{D\overline D}\rightarrow {\cal Z}_{D}{\cal Z}_{\overline D}$ where each factor can be  exactly re-written in terms of a 3-dimensional
effective theory.  We now analyze (\ref{ZDD}) for the SU(2) case following and correcting the arguments in~\cite{DP}.


\bea
{\cal Z}_{D\overline D}[T]&&\equiv \sum_{[K]}\prod_{i_L=1}^{K_L} \prod_{i_M=1}^{K_M} \prod_{i_{\bar L}=1}^{K_{\bar L}} \prod_{i_{\bar M}=1}^{K_{\bar M}}\nonumber\\
&&\times \int\,\frac{fd^3x_{Li_L}}{K_L!}\frac{fd^3x_{Mi_M}}{K_M!}
\frac{fd^3y_{{\bar L}i_{\bar L}}}{K_{\bar L}!}\frac{fd^3y_{{\bar M}i_{\bar M}}}{K_{\bar M}!}\nonumber\\
&&\times {\rm det}(G[x])\,{\rm det}(G[y])\,\,e^{-V_{D\overline D}(x-y)}
\label{SU2}
\eea
with $G[x]$ a $(K_L+K_M)^2$ matrix and $G[y]$ a $(K_{\bar L}+K_{\bar M})^2$ matrix whose explicit form are given in~\cite{DP,DPX}.

\subsection{Classical dyon-antidyon interactions}

\begin{figure}
  \begin{center}
  \includegraphics[width=8cm]{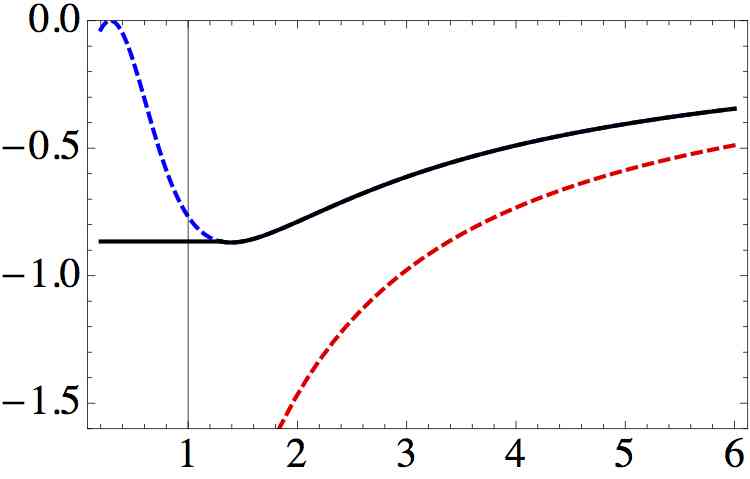}
   \caption{ (Color online) 
   Black solid line is the  SU(2) $D\bar D$ (dimensionless) potential versus the distance $r$ (in units of $1/T$).
   Upper (blue) dashed line is the parameterization proposed in Ref.\cite{LARSEN-SHURYAK}, the lower (red)
    (dashed) line is the Coulomb asymptotics. }
  \label{fig_pot}
  \end{center}
\end{figure}

The explicit form of the Coulomb asymptotic in (\ref{SU2}) for the SU(2) case is

\bea
&&V_{D\overline{D}}(x-y)\rightarrow-\frac {C_D}{\alpha_s\, T}\nonumber\\
&&\times\left(\frac 1{|x_M-y_{\overline{M}}|}+\frac 1{|x_L-y_{\overline{L}}|}-\frac 1{|x_M-y_{\overline{L}}|}-\frac 1{|x_L-y_{\overline{M}}|}
\right)\nonumber\\
\label{DDXX}
\eea
The strength of the Coulomb interaction in (\ref{DDXX}) is $C_D/\alpha_s$ and is of order $1/\alpha_s$. It follows from the asymptotics
of the streamline configuration. In Fig.~\ref{fig_pot} we show the attractive potential for the SU(2) 
streamline configuration in the $M\bar M$ channel~\cite{LARSEN-SHURYAK}. The solid curve is a numerical fit to the 
data given by

\be 
V_{D\bar{D}}(r) \equiv s_{D\bar D}\,V(r)=s_{D\bar D}\,{A v\over g^2}{(r \cdot v-B)^2 \over (r \cdot v)^3+C}  
\label{VMM} 
\ee
with $s_{M\bar M}=-1$ in units of the critical temperature 
$T_c$ and $g$ set to 1 and $A=30.9, B=0.9072, C=15.795$. The dashed line corresponds to the Coulomb
asymptotics

\be
V_{M\bar{M}}(r)\approx -\frac {C_D}{\alpha_s r}
\label{VMMC}
\ee
with $C_D=A/4\pi=2.46$. We recall that in the uncombed $D\bar D$ potential, the 
asymptotic Coulomb interaction corresponds to
$C_D=2$. The attraction in the streamline  is  stronger  asymptotically
owing to the relative combing between the dyons. Fig.~\ref{fig_pot} shows that the $D\bar D$ core is
about $a_{D\bar D}\approx 1/T$.  The second observation is that one should not use the Coulomb asymptotic (the lower dashed curve)
but the actual potential which correctly takes care of the dyons, as extended charged objects rather than
point charges.


Below  the core value of $a_{D\bar D}$, the streamline configuration  annihilates 
into perturbative gluons making the parametrization (\ref{VMM}) arbitrary. Throughout, we will 
parametrize the core by a constant, replacing (\ref{VMM}) by

\be
V_{D\bar D}(r)\equiv s_{D\bar D}\left(V(r)\theta(r-a_{D\bar D})+V(a_{D\bar D})\theta(a_{D\bar D}-r)\right)\nonumber\\
\label{STEP}
\ee
with $s_{M\bar M}=s_{L\bar L}=-1$ (attractive) and $s_{L\bar M}=s_{L\bar M}=+1$ (repulsive).

The ensemble (\ref{SU2}) can be viewed as a 4-component dense and strongly coupled  liquid.  The quantity in the
exponent, known as the classical plasma parameter
\be
\Gamma_{D\bar D}=V(a_{D\bar D})\approx \frac{C_D/\alpha_s a_{D\bar D}}{3T_c}\approx 1
\label{GDD}
\ee
is not small. Its exponent $e^{\Gamma_{D\bar D}}$ can be even large. 
%
This implies 
 that the ``dyonic plasma" we want to study belongs to a class of {\it strongly coupled} plasmas,
 with non-negligible correlations between the particles. So a priori, 
 this problem should be studied by methods more powerful than the 
 usual mean field approximations, such as the Debye-Huckel theory. 
However, we will show below that when the dyonic densities are sufficiently large (and that
implies the overall $T$ of the ensemble to be sufficiently low), the screening mass gets large enough
to put the effective -- screened -- interaction inside the domain in which the analytic Debye-Huckel theory
 becomes justified.

 Furthermore, as we will detail below, the treatment of the repulsive core
is in fact a rather sensitive issue. 
We chose the ``most smooth" version of the potential, shown by the solid curve  in Fig.~\ref{fig_pot}.
Its Fourier transform provides a smooth form factor in momentum space. We note that the actual streamline
was only found for distances $r> a_{D\bar D}\approx 1.2$ (about $4/v$ in the dyon units).
The upper (blue) dashed curve is an example of an arbitrary parameterization discussed in~\cite{LARSEN-SHURYAK}, extending it to smaller
values of $r$. If one uses it, or even  cut off the small $r< a_{D\bar D}$ region 
completely -- the approach known as hard core or excluded volume -- the   Fourier transform of the potential develops large oscillations.
In this case the instability of the  Debye-Huckel theory becomes stronger and its applicability domain shrinks.


The use of (\ref{VMM}) in the repulsive channels $M\bar L$ and $L\bar M$ approximates a smaller repulsion than Coulomb
at shorter distances. A numerical investigation of these channels would be welcome.  Note that both the measure in
(\ref{SU2}) and the asymptotic (\ref{DDXX}) do not include the quantum corrections around the streamline configuration. Both of
which should add more repulsion to the interaction between $D$ and $\bar D$. A leading quantum correction to the 
asymptotic (\ref{DDXX}) follows by analogy from the Coulomb corrections emerging from the $DD$ and $\bar D\bar D$ determinantal
interactions. In our case they are repulsive and amount to the shift 

\be
C_D\rightarrow C_D -\frac{2\alpha_s}\pi +{\cal O}(\alpha_s^2)
\label{SHIFT}
\ee
in the Coulomb constant. The relevance of this correction will be briefly discussed below.

\subsection{Qualitative effect of the one-loop moduli space}

The volume element of the moduli space of
self-dual SU(2) dyons is given by $\sqrt{g_{HK}}\equiv {\rm det}\,G$ with $g_{HK}$
its associated Hyper-Kahler metric~\cite{DP}. 
As we already mentioned in the introduction, the one-loop determinant in the  measure 
(\ref{MEASUREDD}) must be positive  definite for all configurations.
Furthermore, the positivity of all eigenvalues is required, 
since they have the meaning of the volume element in the corresponding subspace.
As noted in \cite{LATTICE},  this is not the case 
 for ensembles with randomly placed  dyons. These  ensembles get denser 
and the positivity condition is only fulfilled for a very small fraction of the configurations.
  

In fact one of the main issues of the dyonic ensembles is the non-trivial character of the 
one-loop  interaction induced by the Diakonov determinant.
 Before we show how this carries to our case through various
fermionization and bosonization and diagrammatic re-summations, it is instructive to provide a qualitative
understanding of the issues using simple explicit examples.

Although it is well known, for completeness let us start with
 the simplest case of two dyons in the SU(2) theory with symmetric 
 holonomy $\nu=\bar\nu=1/2$. Omitting the overall factors,  Diakonov  $2\times 2$ matrix $G$ reads

\bea
G_{2\times 2}[x]\sim 
\begin{pmatrix}
\,&1\pm\frac 1{v x_{12}} & \mp \frac 1 {v x_{12}}  \\
\,&&\,&\,\\
\,& \mp \frac 1{v x_{12}}& 1 \pm \frac 1{ v x_{12}}  
\end{pmatrix}
\label{GMATRIX}
\eea
with $x_{12}\equiv |{\vec x}_{(1)}-{\vec x}_{(2)}|$ the distance between the dyons in units of  $1/v=1/\pi T$. The 
upper signs are for different (ML) dyons, and the lower for similar (MM, LL)   pairs.
The metric-induced potential is thus
$V(x_{12})\equiv -{\rm ln  det}\,G= -{\rm ln}\,(1\pm 2/(v x_{12}))\approx \mp 2/(v x_{12})$ 
is Coulomb-like at large distances. 
(At short distances the induced
potential  is proportional to ${\rm ln}(1/r)$ and not $1/r$.
There is no divergence in the partition function.)

\begin{figure}
  \begin{center}
  \includegraphics[width=7cm]{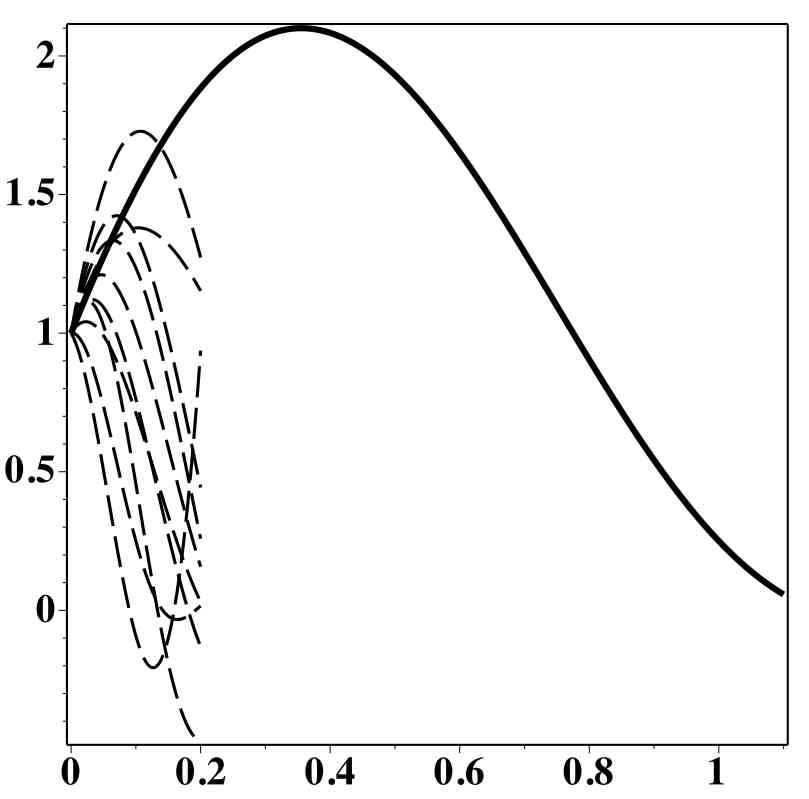}
   \caption{ (Color online)  ${\rm det}\,{G}$ as a function of $\epsilon=1/(\pi aT)$. The dashed lines are for 8 dyons
    randomly placed in a cube of size $a\equiv 1$. The solid line is for correlated dyons in a salt-like or fcc configuration
    also in a unit cube. }
  \label{fig_8dyons}
  \end{center}
\end{figure}

Let us now consider an  ensemble of several ($N=8$) dyons
with $N_M=N_L=4$ and set them randomly in a cube of size $a$. We then evaluate all inter-dyon
distances and calculate  ${\rm det}\, G[x]$ (which is now an $8\times 8$ matrix) 
as a function of the Coulomb parameter $\epsilon=1/(\pi aT)$. For each sampling,
the determinant is a polynomial of $\epsilon$ of degree $N$. The results of 10 random samplings
are displayed in Fig.\ref{fig_8dyons} by the dashed lines. For small $\epsilon$ the determinant deviates
from 1 in a non-uniform way. Some configurations are Coulomb attractive with ${\rm det}\,G>1$, while
some others are repulsive with ${\rm det}\,G<1$ for small $\epsilon$. To first order, they
average to zero for a large number of  charges as there are equal number of positive and negative ones.
At next order, the attraction is to win thanks to the general theorem of second order perturbation theory. 
However, we observe that already for $\epsilon=1/(\pi aT)\sim 0.2$ the repulsive trend is dominant 
and ${\rm det}\,G< 0$ for some samplings. This means that the moduli space of these configurations
vanishes at the corresponding density. This sets an upper limit on the density of random ensembles
of dyons

\be 
n<n_{\rm max}=8\,(0.2\,\pi T)^3\sim 1.98\,T^3
\ee
The lesson: Diakonov determinantal interaction for randomly placed dyons is strongly repulsive, reducing dramatically
the moduli space all the way to zero size for small $\epsilon$. It amounts to a strong effective core 
of order $\alpha_s^0$.

However this is not the end of the story. Let us look at
 the opposite case of a well ordered arrangement of  dyons in the unit box.
 For that we pre-arrange
the 8 dyons of the previous ensemble in a salt-like or fcc configuration on the unit cube, and assess
the corresponding ${\rm det}\,G$. The result is shown in  Fig.~\ref{fig_8dyons} by the solid line. 
While the qualitative trend is the same -- attraction at some interval of densities, changing to repulsion
and then reaching zero at some density -- the value of the maximal density to be reached is
changed by a large factor of about $5^3=125$.  Here is lesson number 2: the moduli space can be made
much larger for the same inter-particle Coulomb strength $\epsilon$, if the correlations between charges are
correctly taken into account.

The overall lesson we get from those examples is the following: Diakonov's original suggestion that
attraction and repulsion would always cancel out is incorrect. Our analysis shows that
ultimately the repulsion
always wins and at some density the volume of the moduli  space always goes to zero. However,
correctly implemented correlations between charges to maximize screening locally, can increase this critical density
by about two orders of magnitude.

\subsection{Fermionization and Bosonization}

Following~\cite{DP} each determinant in (\ref{SU2}) can be fermionized using 4 pairs of ghost fields $\chi^\dagger_{L,M},\chi_{L,M}$ for the dyons
and 4 pairs of ghost fields $\chi^\dagger_{{\bar L},{\bar M}},\chi_{{\bar L},{\bar M}}$ for the anti-dyons. The ensuing Coulomb factors from the determinants are then bosonized using 4 boson fields $v_{L,M},w_{L,M}$ for the dyons and similarly for
the anti-dyons.  The result is a doubling of the 3-dimensional free actions obtained in~\cite{DP}

\bea
&&S_{1F}[\chi,v,w]=-\frac {T}{4\pi}\int d^3x\nonumber\\
&&\left(|\nabla\chi_L|^2+|\nabla\chi_M|^2+\nabla v_L\cdot \nabla w_L+\nabla v_M\cdot \nabla w_M\right)+\nonumber\\
&&\left(|\nabla\chi_{\bar L}|^2+|\nabla\chi_{\bar M}|^2+\nabla v_{\bar L}\cdot \nabla w_{\bar L}+\nabla v_{\bar M}\cdot \nabla w_{\bar M}\right)
\label{FREE1}
\eea
For the interaction part $V_{D\bar D}$, we note that
the pair Coulomb interaction in (\ref{DDXX}) between the dyons and anti-dyons can also be bosonized using
the standard trick~\cite{POLYAKOV}  in terms of $\sigma$ and $b$ fields. Here $\sigma$ and $b$
are the un-Higgsed long range U(1) parts of the original magnetic field $F_{ij}$ and electric potential
$A_4$ (modulo the holonomy) respectively.  As a result each dyon species acquire additional
fugacity factors such that

\be
M:e^{-b-i\sigma}\,\,\,\,\, L:e^{b+i\sigma}\,\,\,\,\, \bar M: e^{-b+i\sigma}\,\,\,\,\, \bar L :e^{b-i\sigma}
 \ee
These assignments are consistent with those suggested in~\cite{UNSAL,TIN} using different arguments.
As a result there is an additional contribution to the free part (\ref{FREE1})

\bea
&&S_{2F}[\sigma, b]=T\int d^3x\, d^3y\\
&&\times\left(b(x)V^{-1}(x-y) b(y)+ \sigma(x)V^{-1}(x-y)\sigma(y)\right)\nonumber
\label{FREE2}
\eea
with $V(r)$ defined in~(\ref{VMM}). The  interaction part is now

\bea
&&S_I[v,w,b,\sigma,\chi]=-\int d^3x \nonumber\\
&&e^{-b+i\sigma}f\left(4\pi v_m+|\chi_M    -\chi_L|^2+v_M-v_L\right)e^{w_M-w_L}+\nonumber\\
&&e^{+b-i\sigma}f\left(4\pi v_l+|\chi_L    -\chi_M|^2+v_L-v_M\right)e^{w_L-w_M}+\nonumber\\
&&e^{-b-i\sigma}f\left(4\pi v_{\bar m}+|\chi_{\bar M}    -\chi_{\bar L}|^2+v_{\bar M}-v_{\bar L}\right)e^{w_{\bar M}-w_{\bar L}}+\nonumber\\
&&e^{+b+i\sigma}f\left(4\pi v_{\bar l}+|\chi_{\bar L}    -\chi_{\bar M}|^2+v_{\bar L}-v_{\bar M}\right)e^{w_{\bar L}-w_{\bar M}}
\label{FREE3}
\eea
In terms of (\ref{FREE1}-\ref{FREE3}) the dyon-anti-dyon partition function (\ref{ZDD}) can be exactly re-written as an interacting
effective field theory in 3-dimensions,

\be
{\cal Z}_{D\overline D}[T]\equiv \int D[\chi]\,D[v]\,D[w]\,D[\sigma]\,D[b]\,e^{-S_{1F}-S_{2F}-S_{I}}
\label{ZDDEFF}
\ee


In the absence of the screening fields $\sigma, b$ (\ref{ZDDEFF}) reduces to the 3-dimensional effective field theory
discussed in~\cite{DP} which was found to be integrable. In the presence of $\sigma, b$ the integrability is lost as the
dyon-anti-dyon screening upsets the hyper-Kahler nature of the moduli space. We will investigate them by linearizing 
the screening effects in the symmetric state.

Since the effective action in (\ref{ZDDEFF}) is linear in the $v_{M,L,\bar M,\bar L}$, the latters   are auxiliary fields that
integrate  into delta-function constraints. However and for convenience, it is best to shift away
the $b,\sigma$ fields from (\ref{FREE3}) through

\be
&&w_M-b+i\sigma\rightarrow w_M\nonumber\\
&&w_{\bar M}-b-i\sigma\rightarrow w_{\bar M}
\label{SHIFT}
\ee
which carries unit Jacobian and no anomalies, and recover them in the pertinent arguments of the delta function constraints as

\bea
&&-\frac{T}{4\pi}\nabla^2w_M+f  e^{w_M-w_L}-f e^{w_L-w_M}=\frac {T}{4\pi}\nabla^2(b-i\sigma)\nonumber\\
&&-\frac{T}{4\pi}\nabla^2w_L+f e^{w_L-w_M}-f e^{w_M-w_L}=0
\label{DELTA}
\ee
and similarly for the anti-dyons.
In~\cite{DP} it was observed that the classical solutions
to (\ref{DELTA}) can be used to integrate the $w^\prime$s in (\ref{ZDDEFF}) to one loop. The resulting bosonic determinant
was shown to cancel against the fermionic determinant after also integrating over the $\chi^\prime$s in (\ref{ZDDEFF}). This
holds for our case as well. However, the presence of $\sigma, b$ makes the additional parts of (\ref{ZDDEFF})  still very
involved in 3 dimensions. 

After inserting the constraints in the 3-dimensional effective action in (\ref{ZDDEFF}), the ground state corresponds
to constant fields because of translational invariance.  Thus, the potential per unit 3-volume $V_3$ following from(\ref{FREE3})
after the shifts (\ref{SHIFT}) is

\bea
-{\cal V}/V_3&&=4\pi f\, \left(v_me^{w_M-w_L}+v_le^{w_L-w_M}\right)\nonumber\\
&&+4\pi f\, \left(v_{\bar m}e^{w_{\bar M}-w_{\bar L}}+v_{\bar l}e^{w_{\bar L}-w_{\bar M}}\right)\nonumber\\
\label{POT}
\eea
Note that if we did not perform the shift (\ref{SHIFT}) then both the potential
(\ref{POT}) and the constraints (\ref{DELTA}) depend on $b$ and $\sigma$ making
the extrema search for ${\cal V}$ more involved. Of course the results should be the same.
For fixed holonomies $v_{m,l}$, the constant $w^\prime$s are real by (\ref{DELTA})
as all right hand sides vanish, and the extrema of (\ref{POT}) occur for

\bea
e^{w_M-w_L}=\pm \sqrt{v_l/v_m}\nonumber\\
e^{w_{\bar M}-w_{\bar L}}=\pm \sqrt{v_{\bar l}/v_{\bar m}}
\label{EXT}
\eea
(\ref{EXT}) is only consistent with  (\ref{DELTA}) if and only if
$v_l=v_m=1/2$ and $v_{\bar l}=v_{\bar m}=-1/2$. That is for confining holonomies or a center
symmetric ground state. However and because of the constraint (\ref{DELTA}) there is no effective potential
for the holonomies in the interacting dyon-anti-dyon liquid. Indeed, by enforcing (\ref{DELTA}) before variation
we have ${\cal V}/V_3=-n_D$,
whatever the $v^\prime$�s. On this point we differ from the arguments and corresponding results
made in~\cite{DP} where the constraints (\ref{DELTA}) were not enforced before the variational derivation
of the holonomy potential. Note that the alternative argument in~\cite{DP}  in favor of the holonomy potential
fixes the number of dyon species $K_i$ to be equal a priori, while (\ref{SU2}) fixes it only on the average.

\section{Linearized Screening Approximation in Center Symmetric State}

For the center symmetric ground state of the 3-dimensional effective theory, we may assess the correction to
the potential ${\cal V}$ to one-loop
in the $b, \sigma$ fields. This is achieved by linearizing the constraints (\ref{DELTA}) around the ground state solutions.
Specifically

\bea
&&\left(-\frac{T}{4\pi}\nabla^2+2 f\right)\,w_M-2fw_L\approx \frac {T}{4\pi}\nabla^2(b-i\sigma)\nonumber\\
&&\left(-\frac{T}{4\pi}\nabla^2+2f \right)\,w_L-2fw_M\approx 0
\label{DELTAL}
\ee
and similarly for the anti-dyons.
The one-loop correction to ${\cal V}$ follows by inserting  (\ref{DELTAL})
in (\ref{ZDDEFF}). The ensuing quadratic contributions before integrations are

\be
S_{1L}=
{\cal V}-4\pi f\int \frac{d^3p}{(2\pi)^3}\frac{(\frac{T}{4\pi} p^2)^2}{(\frac{T}{4\pi}p^2+4f)^2}\left(b(p)^2-\sigma(p)^2\right)\nonumber\\
\label{quad}
\ee
The coefficient of the $b$ field appears tachyonic but is momentum dependent and vanishes at zero momentum.

\subsection{Pressure} 

\begin{figure}
  \begin{center}
 \includegraphics[width=8cm]{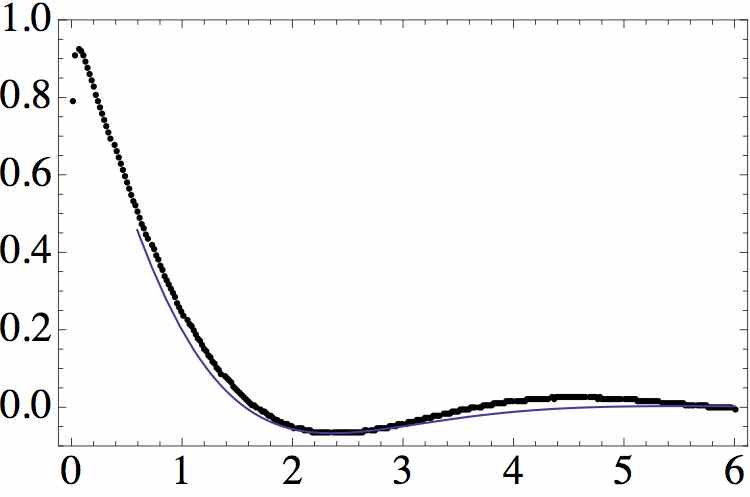}
   \caption{ (Color online) 
   The dots show the form factor, the ratio $V(p)\cdot(p^2/4\pi)$ of  the Fourier transform of (\ref{VMM}) to that of a pure Coulomb law
   versus $p/T$. The thin line is its parameterization. See text.}
  \label{fig_FF}
  \end{center}
\end{figure}

Carrying the Gaussian integration in $b,\sigma$ in (\ref{quad}) yields to one-loop

\bea
{\rm ln} Z_{\rm 1L}/V_3=-{\cal V}-\frac{1}{2}\int \frac{d^3p}{(2\pi)^3}
{\rm ln}\left|1-\frac {V^2(p)}{16}\frac{p^8 M^4}{(p^2+M^2)^4}\right|\nonumber\\
\label{1loop}
\eea
with $V(p)$ the Fourier transform of (\ref{VMM})

\be
V(p)=\frac{4\pi}{p^2}\int_0^\infty dr\,{\rm sin}\,r\,V_{D\bar D}(r/p)
\label{VP}
\ee
and the screening mass  $M=\sqrt{2n_D/T}$ with $|Q^2|=2$ for SU(2).
 In Fig.~\ref{fig_FF} we show the form factor (\ref{VP}) in dots line 
 in units of $T_c$. A simple parametrization is shown in solid line of the form
 
 \be
 V(p)\approx {4\alpha}\,\frac{e^{-p\,a_{D\bar D}}}{p^2}\,\,{\rm cos}(p\,a_{D\bar D})
 \label{FFP}
 \ee
with $\alpha=\pi C_D/\alpha_s$ and a core $a_{D\bar D}\approx 1/T_c$. 
Inserting (\ref{FFP}) into (\ref{1loop}) and setting $\tilde p=p/M$ yield

\bea
{\rm ln} Z_{\rm 1L}/V_3=-{\cal V}-\frac{M^3}{2}\int \frac{d^3{\tilde p}}{(2\pi)^3}
{\rm ln}\left|1-{\tilde\alpha}^2(\tilde p)\frac{{\tilde p}^4}{({\tilde p}^2+1)^4}\right|\nonumber\\
\label{1loopXX}
\eea
with

\be
\tilde\alpha(\tilde p)\equiv \alpha\,e^{-{Ma_{D\bar D}\tilde p}}\,{\rm cos}\,({Ma_{D\bar D}\tilde p})
\label{ALPHAP}
\ee
The dominant contribution to the integral in (\ref{1loop}) comes from the region $\tilde p\approx 1$
for which (\ref{ALPHAP}) can be approximated by $\tilde\alpha(1)\equiv\tilde\alpha$. As a result (\ref{1loop})
can be done approximately by fixing $\tilde\alpha$ and we have the  classical contribution to the pressure

\be
\frac{{\cal P}_{\rm cl}}T\equiv {\rm ln} Z_{\rm 1L}/V_3\approx n_D+\kappa(\tilde\alpha)\frac{M^3}{12\pi}
\label{PRESSURE}
\ee
with

\be
\kappa(\tilde\alpha)=\frac{2+\frac 52 \tilde\alpha+\frac 12 \tilde\alpha^2}{\sqrt{1+\frac{\tilde\alpha}4}}+
\frac{2-\frac 52 \tilde\alpha+\frac 12 \tilde\alpha^2}{\sqrt{1-\frac{\tilde\alpha}4}}-4
\label{KAPPA}
\ee
(\ref{KAPPA}) is seen to vanish for  $\tilde\alpha=0$ or in the absence of $D\bar D$
interactions. Near $T_c$ the screening mass is $M\approx \sigma_E/T_c$ (see below), thus

\be
\tilde\alpha\equiv (\pi C_D/\alpha_s)\,e^{-{Ma_{D\bar D}}}\,{\rm cos}\,({Ma_{D\bar D}})\approx -0.52
\label{ALPHAXX}
\ee
For $|\tilde\alpha|<4$  the 1-loop contribution to the pressure from the charged $D\bar D$ dyons
is real with no dimer or molecular instability. The large core produced by the form factor (\ref{ALPHAP})
is considerably screened by the large dyon density as captured by the large dielectric constant
$1/\kappa(-0.52)\approx 5.26$ in (\ref{PRESSURE}).

The correction in (\ref{PRESSURE}) to the free contribution is a Debye-Huckel 
correction~\cite{DH} (and references therein). A simple but physical way to understand it is to note that a
screened Coulomb charge carries a lower constant  energy

\be
\frac {e^{-M|x|}}{4\pi |x|}\approx \frac 1{4\pi |x|}-\frac{M}{4\pi}+...
\label{DHXX}
\ee
The Debye-Huckel as a mean-field estimate for the pressure follows

\be
\frac{{\cal P}_{DH}}{T}\approx \frac{n_DM}{4\pi T} =\frac{M^3}{8\pi} \rightarrow \frac{M^3}{12\pi}
\label{STANDARD}
\ee
where $n_D=M^2T/2$ is the density of charged particles (see below). The standard Debye-Huckel limiting result 
for a multi-component ionic plasma in 3 spatial dimensions 
is shown on the right-most side of (\ref{STANDARD}).

The correction in (\ref{PRESSURE}) is considerably reduced by the large screening
through the effective dielectric constant played by $1/\kappa(\tilde \alpha)\approx 32/(15{\tilde\alpha}^2)$
for $\tilde\alpha\ll 1$. In particular $1/\kappa (-0.52)\approx 5.26\gg 1$ as noted earlier. It can be recast in the form



\bea
\frac{{\cal P}_{\rm cl}}{T^4}={\tilde n}_D+\frac{\kappa (\tilde\alpha)}{3\pi\sqrt{2}}\,{\tilde n}_D^{\frac 32}
\label{EOS1}
\eea
with ${\tilde n}_D=n_D/T^3$.
Using 
$Ma_{D\bar D}\approx \sigma_E/T_c^2\approx 1/(0.71)^2$ for SU(2) we have 
${\tilde n}_D\approx 1$, so that 
${\cal P}_{\rm cl}/T^4\approx (1+0.01))$. The screening corrections are small of the order of 1\%
thanks to the large dyonic densities.

\begin{figure}
  \begin{center}
  \includegraphics[width=7cm]{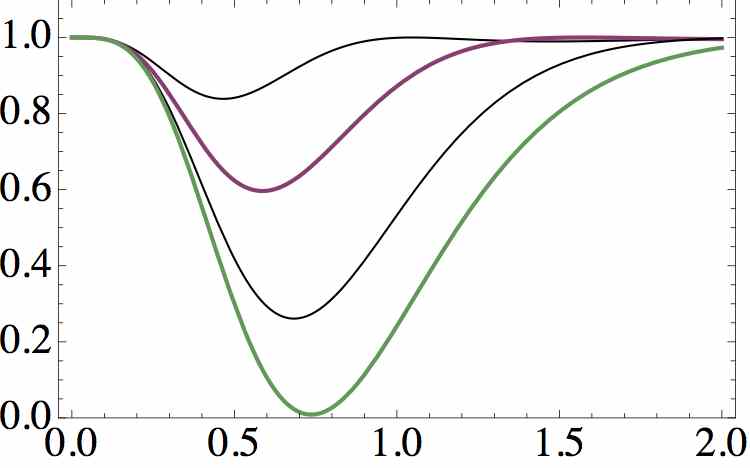}
   \includegraphics[width=8cm]{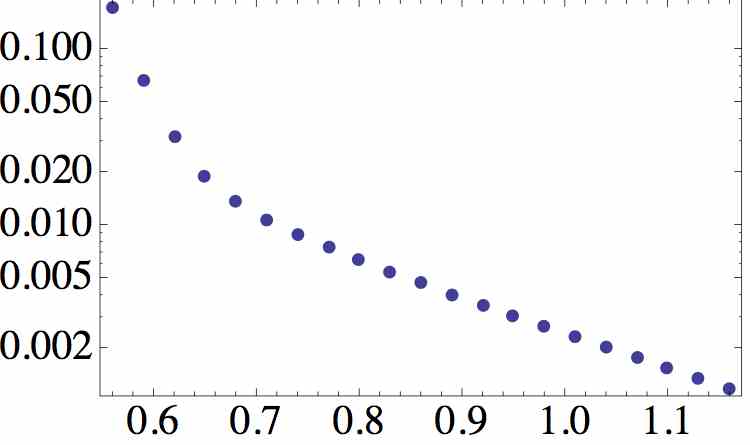}
   \caption{ (Color online) 
   (a) The argument of the logarithm in the last term of (\ref{1loop}) versus the dimensionless momentum $p$, for
   different values of the dimensionless Debye mass $M a_{D\bar D}=1.5,1,0.7,0.56$, top to bottom. As the screening mass decreases
   to its critical value,
   the lower (green) curve touches zero. The smaller values of $M$ leads to a negative argument of the logarithm,  thus
   an instability.
   (b) A semi-logarithmic plot of the integral entering in (\ref{1loop}) as defined in (\ref{INTEGRAND})
   as a function of $M a_{D\bar D}$. The decrease is steady from its maximum at the critical value of the screening mass
   or $Ma_{D\bar D}=0.56$.
     }
  \label{fig_log}
  \end{center}
\end{figure}

The limitations of the Debye-Huckel approximation are readily seen from (\ref{1loop}).
In Fig.~\ref{fig_log}a we plot the argument of the logarithm in the last term of (\ref{1loop}). 
The different curves from top to bottom follow from $Ma_{D\bar D}=1.5, 1, 0.7, 0.56$ respectively.
The smaller the Debye mass $M$ the stronger the dip. For $Ma_{D\bar D}<0.56$, the argument
of the logarithm becomes negative resulting into an $i\pi$ contribution to the pressure and thus an instability.
 This is a clear indication of a well known phenomenon:
  the   Debye-Huckel approximation is in general inapplicable for strongly coupled plasmas,
  and the interaction mediated by the streamline is strong. Only a large enough density
  of dyons, producing sufficiently strong screening, allows for the use of the  Debye-Huckel theory.
  In Fig.~\ref{fig_log}b we show how the total integrated contribution to the free energy 
 changes as a function of the dimensionless Debye mass $Ma_{D\bar D}$

 \be
 (M a_{D\bar D})^3\int_0^\infty dp \,p^2 {\rm ln}\left|1-\frac {V^2(p M)}{16}\frac{p^8 }{(p^2+1)^4}\right| 
 \label{INTEGRAND}
 \ee
The main lesson is that beyond the critical
 value of the screening,  this contribution becomes rapidly very small. This is
 consistent with the analytical estimate above. This justifies the use of the 
 Debye-Huckel mean-field analysis in general,  and the use of the semi-classical expansion 
 in particular.

\subsection{Beyond the Debye-Huckel theory}

The unravelling of the Debye-Huckel  approximation may be due
to  corrections to an interacting Coulomb system, such as
1/ core corrections; 2/ dimer, tetramer and so on many-body interactions. The large core corrections were already
identified and discussed above and yield a substantial reduction in the Debye-Huckel contribution
 near the critical value of $Ma_{D\bar D}\sim 0.56$.

Bound state corrections in the form of electrically charged $L\bar L$ or $M\bar M$ dimers, or electrically neutral
$L\bar L \,M\bar M$ tetramers commonly referred to as instanton-anti-instanton molecules, can bind through the
streamline interaction
(\ref{DDXX}-\ref{VMM}). The combinations $L\bar M$ and $M\bar L$ are repulsive. The binding energy in a dimer
is $\epsilon_{D\bar D}\approx (C_{D}/3)/(\alpha_s a_{D\bar D})=T$. The dimer enhancement is expected to be of order
$e^{\epsilon_{D\bar D}/T}\approx e^{(C_D/3)/\alpha_s}\approx 2.72$ for $T\approx T_c$
using the reduced effective Coulomb coupling.  As we noted earlier, this enhancement becomes
substantially larger at high temperature  as $\alpha_s$ decreases with the onset of dimerization set at about 
$\alpha_{s,{\rm crit}}=\pi (C_D/3)/4\approx 0.67$.  At this coupling which occurs above $T_c$, the Coulomb dimer enhancement factor
is $e^{C_D/3\alpha_{s,{\rm crit}}}\approx 3.57$.

In sum, the dyons and anti-dyons form a
Coulomb liquid with strong short range correlations induced by both the finite cores and bindings.
The liquid supports center symmetry and confines. The deconfinement transition is characterized by
clustering into charged dimers and possibly uncharged and topologically neutral tetramers, forming
mixtures with the restoration of center symmetry. Coulomb mixtures present rich phase diagrams~\cite{FISHER}.

\subsection{Dyonic densities}

(\ref{1loop}) can be readily used to assess the dyon densities $K_M$ and $K_L$ (and similarly for $K_{\bar M}$ and $K_{\bar L}$)
in the center symmetric vacuum with screening
dyons-anti-dyons. For that we need to change $f\rightarrow \sqrt{f_Mf_L}$ and take derivatives
of (\ref{1loop}) with respect
to ${\rm ln}f_{M,L}$ separately and then setting them equal by bulk charge neutrality. The result per species is

\be
K= \frac 14 n_D+\kappa(\tilde\alpha)\frac {M^3}{32\pi}
\label{DENSITY}
\ee
for all dyon and anti-dyon species. 

Each dyon (anti-dyon)  is characterized by an SU(2)  core of size $\rho\approx 1/(2\pi T\,\nu)
\approx 0.33$ fm in  the center symmetric phase with $\nu=1/2$ at $T=1/{\rm fm}$. The Debye length
$\lambda_D=1/M\approx \sqrt{T/2n_D}\approx 0.70\,
{\rm fm}$ is about twice the core size. The classical Coulomb ratio for the ${DD,\bar D\bar D}$ pairs with a core of $2\rho$ is about

\be
\Gamma_{DD,\bar D\bar D}\equiv \frac {1}{2\pi (2\rho)T}\approx \frac {\nu}2=\frac 14
\ee
is small. Recall from (\ref{GDD}) that $\Gamma_{D\bar D}\approx 1$.
The Coulomb ${DD,\bar D\bar D}$  interactions are  quantum and of order $\alpha_s^0$ with strength $1/\pi$
as can be seen by expanding the exponential form of the determinantal interaction in (\ref{GMATRIX}).
The dyon-anti-dyon ensemble is   close to a strongly coupled 4-component Coulomb liquid.  Since the measure for the unlike dyons is exact, it is valid even in the dense configuration. It is only asymptotically exact for like dyons. For the dyons and anti-dyons the streamline is numerically exact
at all separations outside its core. However its corresponding quantum determinant was not calculated. Only a qualitative correction was argued
in  (\ref{SHIFT}).

\subsection{Gluon condensates and susceptibilities}

The topological charge fluctuates locally in this
dyon-anti-dyon model. The topological susceptibility at 1-loop follows from (\ref{1loop}) through the substitution
$f\rightarrow f{\rm cos}(\theta/2)$ both in ${\cal V}$ and also $M\rightarrow M\sqrt{\rm cos(\theta/2)}$.
At finite vacuum angle $\theta$  and in leading order we have

\bea
&&\left<F\tilde F\right>_\theta\equiv - \frac{T}{V_3}\frac{\partial {\rm ln}Z_{1L}}{\partial\theta}=\nonumber\\
&&{\rm sin}(\theta/2)\left(\frac 12 n_DT+\kappa(\tilde\alpha)\frac {M^3T}{16\pi}\sqrt{{\rm cos}(\theta/2)}\right)
\nonumber\\
\label{TOP0}
\eea
Thus the topologicl susceptibility

\bea
\chi_T\equiv \frac {V_3}{T}\left<(F\tilde F)^2\right>_0\approx \left(\frac 12\right)^2 (n_DT)
\label{TOP1}
\ee
in leading order. 
Since the dyons carry half the topological charge (\ref{TOP1}) shows
that the topological fluctuations are  Poissonian to order ${\cal O}(n_D^{3/2})$.
The behavior of $\chi_T/T^4$ versus $T/T_c$ is shown in Fig.~\ref{fig_xit} with 
$n_D$ defined in (\ref{NDX}) below.

  \begin{figure}
  \begin{center}
  \includegraphics[width=7cm]{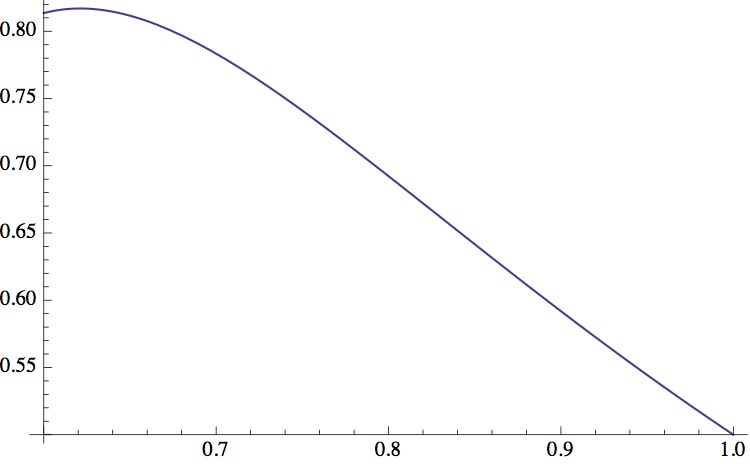}
   \caption{Topological susceptibility in units of $T$ versus $T/T_c$}
     \label{fig_xit}
  \end{center}
\end{figure}

The gluon condensate to 1-loop in the screening approximation follows from

\bea
&&\frac 1{16\pi^2}\left<F^2\right>_0\equiv -\frac {T}{4\pi V_3}\frac{\partial\, {\rm ln} Z_{1L}}{\partial \,{1/\alpha_s}}\nonumber\\
&&\approx -\frac{T}{4\pi}\left(\frac 2{\alpha_s}-\pi\right)\left(n_D+\frac{\kappa(\tilde\alpha)M^3}{8\pi}\right)
\label{TOP2}
\eea
which is non-Poissonian because of the scale anomaly.
The compressibility of the ground state is

\bea
&&\sigma_\chi\equiv \frac{V_3}{T}\left<\left(\frac{F^2}{16\pi^2}\right)^2\right>_c\\
&&\approx \frac{T}{16\pi^2}\left(2\left(n_D+\frac{\kappa M^3}{8\pi}\right)+\left(\frac 2{\alpha_s}-\pi\right)^2
\left(n_D+\frac{3\kappa M^3}{16\pi}\right)\right)\nonumber
\eea
for the connected correlator.

\begin{figure}
  \begin{center}
  \includegraphics[width=7cm]{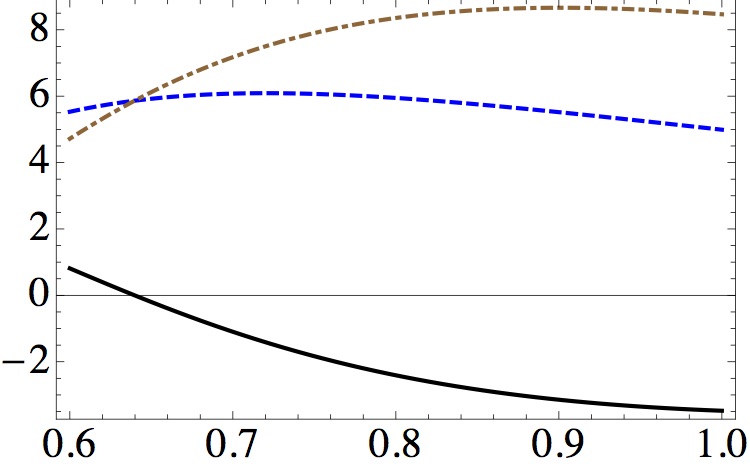}
   \caption{The electric $\left<E^2\right>$ (solid-black), magnetic $\left<B^2\right>$ (dashed-blue) and 
   (Euclidean) energy density $\left<B^2-E^2\right>$  (dot-dashed-brown) 
   in units of $T$ versus $T/T_c$. See text.}
     \label{fig_EEBB}
  \end{center}
\end{figure}

We can use (\ref{PRESSURE}) and (\ref{TOP2}) to extract the electric $\left<E^2\right>_0$ and magnetic  $\left<B^2\right>_0$
condensates in the dyonic ensemble. For that we note that the energy per volume in Euclidean space follows from (\ref{PRESSURE})
through

\be
\frac 1{8\pi}{\left<{B^2-E^2}\right>_0}=T^2\frac{\partial}{\partial T}\frac{P_{\rm cl}}{T}
\label{EEV3}
\ee
The results are

\bea
&&\frac{\left<B^2\right>_0}{4\pi T}=\,\,
\left(+3-(1+\frac {2\alpha_s^\prime}{\alpha_s^2})\left(\frac 1{\alpha_s}-\frac \pi2\right)\right)
\left(n_D+\frac{\kappa M^3}{8\pi}\right)\nonumber\\
&&\frac{\left<E^2\right>_0}{4\pi T}=\left(-3
-(1-\frac{2\alpha_s^\prime}{\alpha_s^2})\left(\frac 1{\alpha_s}-\frac \pi2\right)\right)
\left(n_D+\frac{\kappa M^3}{8\pi}\right)\nonumber\\
\eea
with $\alpha_s^\prime=\partial\alpha_s/\partial\, {\rm ln} T$. In Fig.~\ref{fig_EEBB} we show the behavior of 
the chromo-electric condensate $\left<E^2\right>$ (solid-black), the chromo-magnetic condensate
$\left<B^2\right>$ (dashed-blue) and  the (Euclidean) energy density $\left<B^2-E^2\right>$  (dot-dashed-brown) 
in units of $T$ versus $T/T_c$ in the center symmetric phase. We used the dyon density fixed in
(\ref{NDX}) below.  The chromo-magnetic condensate is about constant in the range of $0.6<T/T_c<1$
while the chromo-electric condensate decreases monotoneously. The condensates are about equal and opposite
near $T_c$ a point supported by the lattice extracted condensates in~\cite{ADAMI}.

\subsection{Electric and magnetic screening masses}

\begin{figure}
\centerline{
\includegraphics[width=6cm]{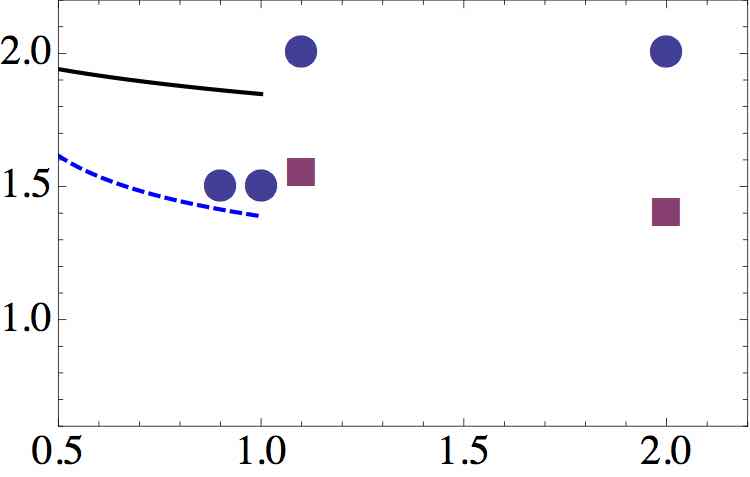}.jpg}
\caption{ (Color online) The electric $M_E/T$ (dashed line) and magnetic $M_M/T$ (solid line) screening masses in (\ref{SCREEME}) versus $T/T_c$.
The points are SU(2) lattice data from ~\cite{BORN} shown for comparison, (blue) circles are  $M_E/T$, (red) squares are $M_M/T$.}
\label{FIGEM}
\end{figure}

The center symmetric phase  of the dyon-anti-dyon liquid screens the long-range U(1) gauge fields left un-Higgsed
by the holonomy $A_4(\infty)/2\pi T= \nu T^3/2$. The electric and magnetic correlations in these Abelian U(1) charges can be
obtained by introducing the U(1) Abelian sources $\eta_{m,e}$ for the magnetic and electric  charge densities

\be
\rho_{m,e}(x)=\sum_iQ_{m,e,i}\,\delta^3( x-{x}_i)
\ee
with $|Q_{m,e}|=1$, and shifting the U(1) fields $\sigma\rightarrow \sigma+\eta_m$ and  $b\rightarrow b+\eta_e$
in the 3-dimensionsl effective action. To 1-loop the generating functional for the charge density correlators is

\be
Z_{1L}[\eta_m,\eta_e]=\int D[\sigma] D[b]\,e^{-S_{2F}[\sigma,b]-S_{1L}[\sigma+\eta_m,b+\eta_e]}
\label{ZEM}
\ee
which is Gaussian in the sources and therefore readily integrated out. Thus

\be
{\rm ln}{Z_{1L}[\eta_m,\eta_e]}=
{-\int \frac{d^3p}{(2\pi)^3}\sum_{i=e,m}\eta_i(p){\bf G}_i(p)\eta_i(-p)}
\ee
with the electric and magnetic density correlators following by variation,

\bea
{\bf G}_{m,e}(p)\equiv \frac 1{V_3}\left<\left|\rho_{m,e}(p)\right|^2 \right>\approx \frac 14
\frac {TM^2p^4}{(p^2+M^2)^2\pm \tilde\alpha \,M^2p^2}\nonumber\\
\label{PAIR}
\eea
The upper sign is for magnetic and the lower sign for electric. 
In x-space, (\ref{PAIR}) can be inverted by Fourier transforms.
The result for the electric correlator in spatial coordinates is


\bea
\label{GEX}
&&-\frac{TM^4}{16\pi |x|}\,e^{-\sqrt{1-\frac {\tilde\alpha}4}\,M|x|}\nonumber\\
&&[
{\rm cos}\left(\frac{\sqrt{\tilde\alpha}}2\,M|x|\right)\,(\tilde\alpha -2) \nonumber\\&&+
{\rm sin}\left(\frac{\sqrt{\tilde\alpha}}2\,M|x|\right)\,\frac{1-2\tilde\alpha+\frac{{\tilde\alpha}^2}2}{\sqrt{\tilde\alpha\,(1-\frac{\tilde\alpha}4)}}]
\eea
The magnetic correlator follows by analytical continuation through the substitution $\tilde\alpha\rightarrow -\tilde\alpha$
in (\ref{GEX}). The electric screening masses $M_{M,E}$ follow from the large distance asymptotics. Using 
our estimate of  $\tilde\alpha\approx -0.52<0$ from the Debye-Huckel analysis above, we have

\bea
&&\frac{{M_E}}{M}\approx \left(\sqrt{1+\frac{|\tilde\alpha|}4}-\frac{\sqrt{|\tilde\alpha|}}2\right)\approx 0.70\nonumber\\
&&\frac{M_M}{M}\approx \left(\sqrt{1-\frac{|\tilde\alpha|}4}  \right)\approx 0.93
\label{SCREEME}
\eea
with $M^2=2n_D/T$.  Also the arguments below  show that
$M=\sigma_E/T$. Combining these two results allow us to fix $C$ in (\ref{NDDD}) above.
Indeed, at $T_c$  the SU(2) lattice results give $T_c/\sqrt{\sigma_E}\approx 0.71$. So (\ref{NDDD}) now reads

\be
\frac{n_D}{T^3}\approx 2\,\frac{\alpha^2_s(T_c)}{\alpha^2_s(T)}\,e^{\frac \pi{\alpha_s(T_c)} -\frac \pi{\alpha_s(T)}}
\label{NDX}
\ee
which gives $M_{E}\approx 1.4\,T_c$ 
and $M_M\approx 1.8\,T_c$, both of  which are remarkably close to the reported SU(2) lattice results
in the vicinity of the critical temperature~\cite{BORN}. In Fig.~\ref{FIGEM}
we display the results (\ref{SCREEME})  for $M_{E,M}/T$ in the range $(0.5-1)\,T_c$ 
versus $T/T_c$. The points at $T>T_c$ are shown for comparison. We
note that the electric mass drops down at $T_c$. In the region we study $M_M>M_E$, while above $T_c$, in a more familiar
QGP region, $M_M<M_E$. This switching of the magnitude of the two screening masses
is better documented in lattice works with the $SU(3)$ gauge group. It has a simple explanation in our case.
Since at $T>T_c$ the dyon density drops it follows that $M$ decreases as well. As a result, the form factor in Fig.~\ref{fig_FF}
is probed at smaller momentum $p\approx M$ (larger distances) making $\tilde\alpha(p\approx M)$ in (\ref{ALPHAP}) 
switch from negative ($T<T_C$) to positive ($T>T_C$).
From (\ref{PAIR}) it follows that the expressions for $M_{E,M}$ in (\ref{SCREEME}) are now switched with
$M_M$ lighter than $M_E$. A simple estimate of the critical temperature at the crossing follows from the vanishing
of (\ref{ALPHAXX}) or $Ma_{D\bar D}=\pi/2$. This translates to a critical dyon density 
$n_D^C\approx \pi^2T_c^3/8$ which is consistent
with our estimate of $T_c$ below (see (\ref{TCC})).

Finally, we note  that 
the value of $\alpha_s(T_c)\approx 0.84$ extracted from the cooled caloron data in~\cite{SHURYAK}
is also consistent with the reported value from bulk thermodynamics in~\cite{ALPHA}. 
In the dyon-anti-dyon Coulomb liquid  the correlators are modified
 at intermediate distances as we now detail in terms of the static structure factors.

\subsection{Static structure factors}

\begin{figure}
\centerline{
\includegraphics[width=6cm]{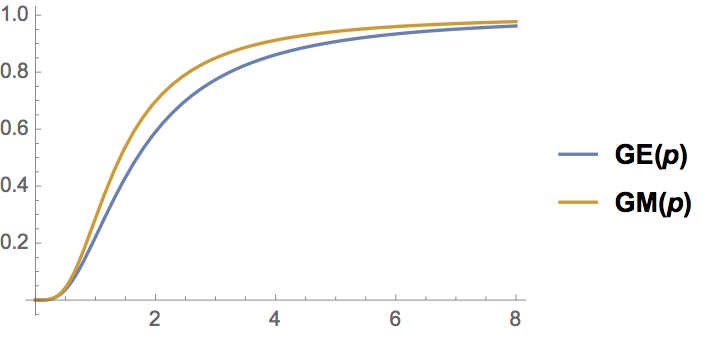}}
\caption{ (Color online) The electric and magnetic  structure factors (\ref{PAIR}) as a function of $p/M$.}
\label{FIGME}
\end{figure}

The charged structure factor between pair of magnetic or electric charges is
 (\ref{PAIR}) which can be re-written as

\be
{\bf G}_{M,E}(p)\equiv \left<\frac 1{N_{m,e}}\left| \sum_{j=1}^{N_{m,e}}Q_{m,e,j}e^{ik\cdot x_j}\right|^2\right>
\ee
Thus

\be
{\bf G}_{M,E}(p)=\frac {{\bf G}_{m,e}(p)}{n_D/2}
\equiv \frac{\tilde p^4}{(\tilde p^2+1)^2\pm \tilde{\alpha}{\tilde p}^2}
\label{PAIRME}
\ee
with $\tilde p=p/M$.
We note that the pre-factor in (\ref{PAIRME}) involves  two static  electric or
magnetic exchanges with an identical screening mass $M$. 
The charged  structure factors vanish as ${\bf G}_{M,E}(p)\approx {\tilde p}^4$.
For large momenta or $\tilde p \gg 1$  both structure factors asymptote one from below
as shown in Fig.~\ref{FIGME}. The magnetic hole is slightly smaller than the electric one
around the same pairs. The absence of oscillations in the structure factor,
 is a consequence of our linearized approximation.

To characterize further the 4-component plasma of dyons and anti-dyons we define the scalar
static pair correlation function

\be
{\bf G}_S(x)=\left<\frac 1N\sum^N_{i\neq j}\delta^3(x+x_i-x_j)\right>
\label{PAIRS}
\ee
normalized to the total number of particles $N$. (\ref{PAIRS}) defines
the probability to find two particles a distance $|x|$ apart. Its Fourier transform

\be
{\bf G}_S(p)=\left<\frac 1N\left| \sum_{j=1}^{N} e^{ip\cdot x_j}\right|^2\right>
\label{PAIRSK}
\ee
is the scalar structure factor.

(\ref{PAIRSK})  can be evaluated by switching $f\rightarrow f+\delta f (x_i)$ in (\ref{SU2}) and then
linearizing the resulting effective action around the mean-density. Specifically, we can re-write the
linearized constraint (\ref{DELTAL}) formally as

\be
(w_M-w_L)=\frac{1}{\Delta_0+4\delta f}\left(\frac{T\nabla^2}{4\pi}\right)(b-i\sigma)
\label{PERT}
\ee
with $\Delta_0=-T\nabla^2/4\pi+4f$ and use perturbation theory to expand the denominator in
(\ref{PERT}) to order ${\cal O}(\delta f^3)$.  
The result can be formally written as

\be
(w_M-w_L)(p)=\int\frac{d^3k}{(2\pi)^3}G(p,k)(b-i\sigma)(k)
\label{PERT1}
\ee
Inserting (\ref{PERT1}) into the potential (\ref{POT}) 
yields a quadratic action in $b$ and $\sigma$. Integrating over the latters yields the 1-loop determinant or
the effective action for $\delta\equiv \delta f/f$. Specifically

\be
{\rm det}(1+{\bf S}[\delta ])=e^{{\rm Tr ln}(1+{\bf S}[\delta ])}\approx e^{{\rm Tr}\, {\bf S}[\delta ]}
\ee
with the quadratic effective action for the scalar fluctuations as

\be
{\rm Tr}\, {\bf S}[\delta ]={-\int \frac{d^3p}{(2\pi)^3}\,\delta (p){\bf G}^{-1} (p)\delta (-p)}
\label{TSCA}
\ee
with to order ${\cal O}({\tilde\alpha}^3)$

\bea
{\bf G}^{-1}(p)&&\approx n_D +
4\,{\tilde\alpha}^2M^8\int \frac{d^3k}{(2\pi)^3}\nonumber\\
&&[
\,\,\frac{k^4}{(k^2+M^2)^4}\frac{1}{((k+p)^2+M^2)^2}\nonumber\\
&&+\frac{2k^2}{(k^2+M^2)^3}\frac{1}{((k+p)^2+M^2)^2}\,\,]
 \label{GSCA}
\eea
The $n_D$ contribution in (\ref{GSCA}) follows from the expansion of the  leading contribution ${\cal V}$
using  arguments similar to those used for the derivation of the dyonic densities above.

The scalar structure factor follows from (\ref{GSCA}) through the normalization

\be
{\bf G}_S(p)\equiv \frac {n_D}{V_3}\left<\left|\delta (p)\right|^2\right>=n_D\,{\bf G}(p)
\label{NSCA}
\ee
We note that the small momentum fluctuations in $\delta f$ couple 
to the sound-like modes. Specifically,

\be
{\bf G}_S(p) \approx \frac{p^2}{c_s^2p^2}
\ee
is dominated by a massless pole at zero momentum
with

\be
c_s^2\approx 1 +8\,{\tilde\alpha}^2\left(\frac MT\right)
\int\frac{d^3k}{(2\pi)^3}\left[ \frac {k^4}{(k^2+1)^6}+\frac{2k^2}{(k^2+1)^5}\right]\nonumber\\
\label{CS2}
\ee
Alternatively from the pressure (\ref{PRESSURE}) we expect

\be
c_s^2\equiv \frac{\partial {\cal P}_{\rm cl}}{T\partial n_D}\approx 
1 +\frac{\kappa(\tilde\alpha)}{4\pi}\left(\frac MT\right)
\ee
with $\kappa(\tilde\alpha)\approx 15\,{\tilde\alpha}^2/32$ in leading order and in total
agreement with (\ref{CS2}). Also, at large momentum (\ref{NSCA}) asymptotes
${\bf G}_S(\infty)=1$. The slight super-luminal character of (\ref{CS2}) reflects
on the fact that dyons are in essence Euclidean configurations with no physical
particle realization.

\subsection{Estimate of the critical $T_c$}
\label{sec_deconf}

The total thermodynamical pressure of the dyon-anti-dyon liquid  consists of the classical and non-perturbative
contribution (\ref{EOS1}) plus the perturbative  holonomy potential known as 
Gross-Pisarski-Yaffe-Weiss (GPYW) potential
\cite{WEISS}, plus the purely perturbative
black-body contribution (ignoring the higher order ${\cal O}(\alpha_s)$ quantum corrections). Specifically $(N_c=2$)

\bea
\frac{{\cal P}_{\rm tot}}{T^4}&&\approx  {\tilde n}_D+\frac{\kappa (\tilde\alpha)}{{3\pi\sqrt{2}}}\,{\tilde n}_D^{\frac 32}
\nonumber\\
&&-\frac{\pi^2}{45}\left(N_c^2-\frac 1{N_c^2}\right)+\frac{\pi^2}{45}\left(N_c^2-1\right)
\label{EOS2}
\eea
The Debye-Huckel contribution is of order $N_c^{3}$, while the leading
classical contribution is of order $N_c^2$. So screening and large $N_c$ are not commutative.
For the SU(2) case of interest, the transition temperature $T_c$
from the disordered phase ($\nu=1/2$) to the ordered phase  ($\nu=0$)
occurs when the first three contributions in (\ref{EOS2}) cancel out.  Thus

\be
{\tilde n}_D+\frac{\kappa (\tilde\alpha)}{{3\pi\sqrt{2}}}\,{\tilde n}_D^{\frac 32}
\approx \frac{\pi^2}{12}
\label{TCC}
\ee
For $\kappa(-0.52)\approx 0.19$, the critical density is 
$n^C_{D}\approx {\pi^2 T_c^3}/{12}\approx 0.88$.
Since the SU(2) electric string tension 
is $\sigma_E=TM=\sqrt{2n_DT}$ (see below), 
it follows that $T_c/\sqrt{\sigma}=6^{\frac 14}/\sqrt{\pi}\approx 0.88$ which is somehow
larger than the SU(2) lattice result $T_c/\sqrt{\sigma_E}=0.71$~\cite{TEPER}.  

\section{Polyakov lines}

To probe the confining nature of the dyon-anti-dyon liquid  in the 3-dimensional effective theory we will compute explicitly
the expectation of a heavy quark through the traced Polyakov line and the corelator of a heavy quark-anti-quark pair through
the correlator of the traced Polyakov line and its conjugate at fixed spatial separation. The insertion of these charges in the
dyon-anti-dyon liquid modifies the ground state through solitonic solutions around these sources.

In this section we present
a new derivation of the pertinent solitonic equations for the SU(2) case
that makes explicit use of the presence of the long range U(1) $b$ and $\sigma$
fields. In the linearized screening approximation, we show that the solitonic equations for the heavy source probes are in
agreement with those established in~\cite{DP} using different arguments.

\subsection{$\left<L\right>$}
\label{LSECTION}

In the SU(2) case the Polyakov line consists of inserting a heavy quark whose free energy consists of its Coulomb interactions
with all the Coulomb charged dyons and anti-dyons. Specifically, the traced Polyakov line before averaging is

\bea
\label{L1X}
&&L(x_1)=\\
&&e^{2\pi i\mu_M+\frac{i}{2T}\sum_{i}\left(\frac{1}{|x_1-x_{Mi}|}+ \frac{1}{|x_1-x_{\bar Mi}|}-\frac{1}{|x_1-x_{Li}|}-\frac{1}{|x_1-x_{\bar Li}|}\right)}+\nonumber\\
&&e^{{2\pi i\mu_L}+\frac  i{2T}\sum_{i}\left(\frac{1}{|x_1-x_{Li}|}+ \frac{1}{|x_1-x_{\bar Li}|}-\frac{1}{|x_1-x_{Mi}|}-\frac{1}{|x_1-x_{\bar Mi}|}\right)}
\nonumber
\eea
with $\mu_L-\mu_M=v_m$.
When averaging using the ensemble (\ref{MEASUREDD}) it is clear that each of the contributors to the string of factors in (\ref{L1X})
will match its analogue from the measure and re-exponentiate.
For instance the first contribution  in (\ref{L1X})  re-exponentiates through the substitution

\bea
e^{-b\pm i\sigma} \rightarrow &&e^{-b\pm i\sigma}e^{\frac{i}{2T|x_1-x|}}
\eea
The extra Coulomb factors can be re-defined away by shifting 

\be
b\rightarrow b+\frac{i}{2T|x_1-x|}
\ee
thereby changing the  constraint equation (\ref{DELTA}) to

\bea
&&-\frac{T}{4\pi}\nabla^2w_M+f(e^{w_M-w_L}-e^{w_L-w_M})\nonumber\\
&&=\frac{T}{4\pi}\nabla^2(b-i\sigma)+\frac{i}{2}\delta^3(x-x_1)\nonumber\\
&&-\frac{T}{4\pi}\nabla^2 w_L+f(e^{w_L-w_M}-e^{w_M-w_L})=0
\label{SOLM}
\eea
and similarly for the second contribution in (\ref{L1X}) with $L\leftrightarrow M$.
The effect of the first contribution in the Polyakov line (\ref{L1X}) is to add a source
term to the constraint equation for $w_{M}$. It is in agreement  with~\cite{DP}
after setting $b=\sigma=0$. (\ref{SOLM}) is a Poisson-Boltzmann type equation.
It is also referred to as an elliptic and periodic Toda lattice~\cite{DP,TODA}.
The solution  is a local Debye-like cloud around the inserted heavy quark

\be
(w_M-w_L)(x)\approx  \frac{2\pi i}{T}\int \frac{d^3p}{(2\pi)^3}\frac{e^{ip\cdot (x-x_1)}}{p^2+M^2}
\ee
This  causes almost no change in the vacuum holonomies $v_{m,l}$. Thus, after the shift

\be
\left<L(x_1)\right>\approx e^{i2\pi\mu_M}+e^{i2\pi\mu_L}=0
\label{L00}
\ee

\subsection{$\left<LL^\dagger\right>$}

The preceding analysis can also be applied to the correlator of two heavy quarks through $LL^\dagger$
which consists now of 4 contributions before averaging

\be
L (x_1)L^\dagger(x_2)=\sum_{m,n=M,L}e^{2\pi i (\mu_{m}-\mu_{n})}e^{\frac{i}{2T}(F_{m}(x_1)-F_{n} (x_2))}\nonumber\\
\label{LL1}
\ee
with the pertinent Coulomb free energies $F_{m}(x_{1,2})$ following from (\ref{L1X}).
When averaged over the measure (\ref{MEASUREDD}), each of the factors in (\ref{LL1})
can be matched  with its analogue in the measure.  The preceding observations
show that the Coulomb factors in the probing correlator can be paired with

\bea
e^{-b\pm i\sigma}\rightarrow e^{-b\pm i\sigma}e^{\frac{i}{2T|x-x_1|}-\frac{i}{2T|x-x_2|}}
\label{LL1X}
\eea
A rerun of the preceding arguments shows that the constraint equations acquire now two source contributions, one for each of the
heavy quark inserted

\bea
&&-\frac{T}{4\pi}\nabla ^2w_M+f(e^{w_M-w_L}-e^{w_L-w_M})\nonumber \\ 
&&=\frac{T}{4\pi}\nabla^2(b-i\sigma)+\frac{i}{2} [\delta^3(x-x_1)-\delta^3 (x-x_2)] \nonumber\\
&&-\frac{T}{4\pi}\nabla^2w_L+f(e^{w_L-w_M}-e^{w_M-w_L})=0
\label{LL2}
\eea
Since $\nabla^21/|x-x_2|=-4\pi\delta^3(x-x_2)$,
we can symmetrize (\ref{LL2}) by shifting $\delta^3(x-x_2)$ from the first to the second equation
through

\be
w_{M,L}\rightarrow w_{M,L}+\frac{i/T}{2|x-x_2|}
\label{LL2X}
\ee
with unit Jacobian. The symmetrized (\ref{LL2}-\ref{LL2X}) equations are in agreement with those established in
\cite{DP} for the SU(2) case after setting $b=\sigma=0$. In this case the solution is peaked around the heavy quark sources

\be
(w_M-w_L)(x)\approx  \frac{2\pi i}{T}\int \frac{d^3p}{(2\pi)^3}\frac{e^{ip\cdot (x-x_1)}-e^{ip\cdot (x-x_2)}}{p^2+M^2}
\ee
Inserting this back in the expectation value of the
correlator (\ref{LL1}) yields asymptotically

\be
\left<L (x_1)L^\dagger(x_2)\right>\approx e^{-M|x_1-x_2|}
\label{LLCONF}
\ee
in the 3-dimensional effective theory in agreement with the result in~\cite{DP}. In 4-dimensions 
(\ref{LLCONF}) translates to confinement of the
electric charges with the electric string tension $\sigma_E=MT$.
The additional Coulomb screening in (\ref{MEASUREDD}) does not affect the asymptotics
of the linearly rising heavy quark potential to leading order. 
The dyon-anti-dyon Coulomb liquid still electrically confines in the center  symmetric  phase.

\section{Single-Winding Wilson loop}
\label{WILSONSECTION}

To study the large spatial Wilson loops we use the same observations made above in the
presence of the U(1) fields $\sigma$ and $b$. As an observable the traced spatial Wilson loop
of area $S$ supported by the spatial contour $\partial S=C$ reads

\be
{\rm Tr}\,W(C)=  e^{i\int_{S} B_{+}\cdot dS} +e^{i\int_{S} B_-\cdot dS}
\label{WC1}
\ee
and sources the static magnetic field

\be
B_{\pm\,\mu}= \pm\sum_i Q_{i} \frac{(x-x_i)_{\mu}}{|x-x_i|^3}
\label{WC2}
\ee
When averaged using (\ref{MEASUREDD}), the spatial Wilson loop (\ref{WC1}) modifies the
 additional U(1) fugacity factors in the dyon sector. Their contribution follows again by shifting
$b\mp i\sigma\rightarrow b\mp i(\sigma-\eta_{\pm})$ in the constraint equations  with


\be
\eta_{\pm}(x)=\pm \int_{S} dS_{y}\cdot\frac{x-y}{2|x-y|}
\ee
As a result,  (\ref{DELTA}) in the presence of (\ref{WC1}) are now modified to read

\bea
&&-\frac{T}{4\pi}\nabla^2w_M+f(e^{w_M-w_L}-e^{w_L-w_M})=\frac{iT}{4\pi} \nabla^2\eta_+(x) \nonumber
\\&&-\frac{T}{4\pi}\nabla^2w_L+f(e^{w_L-w_M}-e^{w_M-w_M})=0
\label{WC3}
\ee
for the first contribution and similarly
for the second contribution in (\ref{WC1}) with $\eta_+\rightarrow\eta_- $.  After choosing the spatial Wilson loop to lay in the x-y plane
through $\nabla^2\eta_{\pm}=\pm 4\pi\delta^\prime(z)$,
the results (\ref{WC3}) are in agreement with those derived in \cite{DP} for the SU(2) case
but without the long range U(1) $\sigma$ and $b$ fields in the leading order approximation. Thus
$\left<{\rm Tr}\,W(C)\right>\approx e^{-\sigma_M \,S}$ is saturated by the pinned soliton,
with the magnetic string tension $\sigma_M=\sigma_E=MT$.
This result is expected from the equality of the
electric and magnetic masses in (\ref{SCREEME}).

A simple understanding of this result is as follows: while a heavy quark sources an electric field, a
large spatial Wilson loop sources a magnetic field by Ampere$^\prime$s law
which is classically composed of all the magnetic poles fluxing $S$
as is explicit in (\ref{WC2}). The typical contribution to  (\ref{WC1})
for a planar surface in the xy-plane is then

\be
\left<e^{i\int_{S} B\cdot dS}\right>\approx e^{-\frac S2\int_S \left<B_z(x,y)B_z(0,0)\right>dS}
\ee
by keeping only the first cumulant in the average and using translational invariance  for
large $S$. In this limit, $S$ acts as a uniformly charged magnetic sheet made of magnetic
dyons classically, so that

\be
\left<B_z(x,y)B_z(0,0)\right>\approx \left<\left(\frac{Q_M}{S}\right)^2\right>\approx \frac{\left<Q_M\right>}{S^2}
\ee
where the variance in the magnetic charge is assumed Poissonian. The magnetic charge density per unit
4-volume is $(TM)^2/2$. The typical magnetic charge per unit area is  then about its square root or
$\left<Q_M\right>/S\approx TM$.  Thus

\be
\left<e^{i\int_{S} B\cdot dS}\right>\approx e^{-\frac 12 MT\,S}
\ee
which is the expected behavior up to a factor of order one in the string tension.

\section{Double-Winding Wilson Loop}

Recently it was pointed out in~\cite{GREEN} that a co-planar and double winding Wilson loop in the SU(2) 
pure gauge theory version of the model discussed by Diakonov and Petrov~\cite{DP} shows an exponentl
fall-off with the sum of the areas. In contrast lattice SU(2) simulations appear to show an exponential fall off
with the difference of the areas. The main observation in~\cite{GREEN} is that the solitonic configuration
contributing to the single-winding spatial Wilson loop as for instance from our linearized version with
$b=\sigma=0$ in (\ref{WC3}),  factors out in the the double-winding and co-planar Wilson loop.

For two
identical loops with $C_1=C_2=C$, we have
the formal SU(2) identities~\cite{GREEN1} (and references therein)

\bea
\left({\rm Tr}\, W(C)\right)^2 =&&{\rm Tr}_S\left(W(C)\right)+{\rm Tr}_A\left(W(C)\right)
\nonumber\\
{\rm Tr}\left(W(C)^2\right) =&&{\rm Tr}_S\left(W(C)\right)-{\rm Tr}_A\left(W(C)\right)
\label{DW1}
\ee
The simple trace ${\rm Tr}$ is carried over the fundamental representation of N-ality
$k=1$ as in (\ref{WC1}), and ${\rm Tr}_{S,A}$ are carried over the symmetric and anti-symmetric of N-ality $k=2$ (modulo 2)
 representations of SU(2) respectively. The identities (\ref{DW1}) are commensurate with the Young-Tableau
 decomposition. In the dyonic plasma considered here, the k-string tensions $\sigma_k$
 in the linearized plasma approximation are
 identical to those derived in~\cite{DP} with $\sigma_k/\sigma_1={\rm sin}\,{k\pi/2}$ for SU(2)
 with $\sigma_1=\sigma_E$. 
 For $k=2$ we have $\sigma_2=0$ and the second identity in (\ref{DW1}) implies for large loops

 \bea
\left<{\rm Tr}\left(W(C)^2\right)\right> =\left<{\rm Tr}_S\left(W(C)\right)\right> -1
\label{DW1}
\eea
We have set all self-energies to zero for simplicity as they depend on the subtraction procedure.
(\ref{DW1}) is consistent with the doubly traced Wilson loop as dominated by the $k=2$ modulo
2 colorless di-quark-like $(qq)$ or baryon-like configuration in SU(2). In the dyonic plasma, the  double
Wilson loop with $C_1=C_2$ is a bound colorless state with zero-size that is strongly
correlated within the dyons cores and therefore is consistent with the arguments presented
in~\cite{GREEN}.

For largely separated loops $C_{1,2}$ of arbitrary sizes but still lying in the spatial directions,
clearly 

\be
\left<{\rm Tr}(W(C_1)W(C_2))\right>\approx e^{-\sigma_E(A_1+A_2)}
\ee
for $(A_1+A_2)<A_{12}$ where $A_{1,2}$ are the planar areas supported by $C_{1,2}$ separately,
and $A_{12}$ is the minimal area with boundaries $C_1$ and $C_2$.
The main issue  is what happens for the same doubly wound SU(2)
spatial Wilson loops when $A_{12}<(A_1+A_2)\,? $  Here we note that  $L\bar L$ and $M\bar M$
dimers carrying $(-2,0)$ and $(+2,0)$ ${\rm (electric, magnetic)}$ charge assignments could 
cluster around the probe $qq$ (baryon) and $\bar q \bar q$ (anti-baryon) configurations respectively,
to form neutral molecular bound states with masses that scale with $A_{12}$  instead of $(A_1+A_2)$.
They are commensurate with the 
massive off-diagonal and charged gluons Higgsed by the holonomy and dropped in the dyon liquid
analysis. These configurations were not  retained in~\cite{DP}.

\section{t$^\prime$ Hooft Loops}

In an important study of the nature of confinement in gauge theories, t$^\prime$ Hooft~\cite{HLOOP} has introduced
the concept of a disorder operator or t$^\prime$ Hooft  loop to quantify confinement in the Hilbert space
of gauge configurations. The t$^\prime$ Hooft  loop is a canonical operator much like the Wilson loop. In
a Lorentz invariant confining vacuum, t$^\prime$ Hooft has argued that the temporal
Wilson loop and the t$^\prime$ 
Hooft  loop cannot exhibit an area law simultaneously. The temporal Wilson loop obeys an area law, while the 
t$^\prime$ Hooft  loop obeys a perimeter law.

Physically, the Wilson loop corresponds to a color 
charge in the fundamental representation running around a closed loop  and  measuring the 
the chromo-magnetic flux across the loop. The t$^\prime$ Hooft loop corresponds to a dual charge 
 in the center of the gauge group running around a closed loop and 
measuring the   chromo-electric  flux  across the loop. The t$^\prime$ Hooft loop is  the dual of the
Wilson loop. 

In the temperature range $0.5\,T_c<T<T_c$  confinement is still at work and we expect the 
temporal Wilson and  t$^\prime$ Hooft loops to exhibit behaviors similar to those in the vacuum state. 
In section~\ref{WILSONSECTION} we have explicitly checked that the closed spatial Wilson loop
obeys an area law. The temporal Wilson loop is  not amenable to our dimensionally reduced 
and Euclideanized effective field theory.

The t$^\prime$ Hooft loop $V(C)$ enforces a gauge transformation $\Omega_C$ which is singular
on a closed curve $C$. If a curve $C^\prime$ winds $n_{CC^\prime}$ times around $C$  then

\bea
V^\dagger (C)W(C^\prime)V(C)=e^{i\frac {2\pi}{N_c}n_{CC^\prime}}W(C^\prime)
\label{TH1}
\eea
$V(C)$ amounts to a multi-valued gauge transformation on the loop $C$,

\be
\Omega_C (\theta = 2\pi )=e^{i\frac {2\pi}{N_c} n_{CC^\prime}}\Omega_C (\theta =0)
\label{TH0}
\ee
with $\theta$ an affine parameter along $C$.  A simple choice is

\be
\Omega_C(x)=e^{i\frac {2\pi}{N_c}Q\varphi_C(x)}
\label{TH2}
\ee
where $\varphi_C(x)$ is a multi-valued scalar potential for the magnetic field $\vec B_C$ generated by a loop
of current $\vec j_C$ running along $C$, and $Q=(1, 1, ..., -N_c+1)$ a Cartan generator of SU(N$_c$). An alternative
construction using a discontinuous solid angle
was discussed in~\cite{KOSTAS,HUGO}. The effects of (\ref{TH2}) on an Abelianized Wilson loop is

\be
\Omega_C^\dagger \left(e^{i\int_{C^\prime}\,ds\cdot A}\right) \Omega_C
=e^{i\int_{C^\prime} ds\cdot (A-\frac {2\pi}{N_c}QB_C)}
\label{TH3}
\ee
with $\vec B_C=-\vec\nabla\varphi_C$. Note that since $\varphi_C$ is {multivalued} we have
$\vec\nabla\times \vec B_C=4\pi \vec j_C$. If we normalize  the loop current $\vec j_C$ such that

\be
\int_{C^\prime}ds\cdot B_C=4\pi\int_{A(C^\prime)}dS\cdot j_C=-n_{CC^\prime}
\label{TH4}
\ee
then (\ref{TH3}) reduces to 

\be
\Omega_C^\dagger \left(e^{i\int_{C^\prime}\,ds\cdot A}\right) \Omega_C 
=e^{i\frac {2\pi}{N_c}n_{CC^\prime}}e^{i\int_{C^\prime}\,ds\cdot A}
\label{TH5}
\ee
In the space of gauge configurations, the gauge transformation $\Omega_C$ is inforced through

\be
V(C)=e^{i\frac {2\pi}{gN_c} \int d^3x {\rm Tr}\left( E_i D_i (Q\varphi_C)\right)}
\label{TH6}
\ee
For SU(2) we have

\be
V(C)=e^{i\frac{2\pi}{g} \int d^3x \vec{E^3}\cdot \vec B_C}\rightarrow e^{-\frac{2\pi}{g} \int d^3x \vec{E^3}\cdot \vec B_C}
\label{TH7}
\ee
where the latter substitution 
$E\rightarrow iE$ follows in Euclidean space. With this in mind, the expectation value
of the t$^\prime$ Hooft loop  in the dyonic ensemble  involves a string of sources to be inserted in
(\ref{SU2}). In leading order

\be
V(C)\rightarrow &&\prod_{i=1}^{N+\bar N}
e^{\frac{2\pi}{g}\int d^3x\,B_C\cdot\nabla\frac {Q_{Ei}}{|x-x_i|}}\nonumber\\
=&&\prod_{i=1}^{N+\bar N}
e^{-\frac{2\pi}{g}\int d^3x\,\nabla\cdot B_C\frac {Q_{Ei}}{|x-x_i|}}=1
\ee
Thus $\left<V(C)\right>=1$ modulo ${\cal O}(\alpha_s)$ Coulomb-like self-energy corrections which are
perimeter-like in general.

Finally, the Polyakov line as a Wilson loop around the periodic temporal direction
has a dual Polyakov loop with a  dual magnetic charge in the center. In the confined phase, 
the temporal component of the gauge field $A_4$ asymptotes fixed electric-type holonomies, while its
dual $\tilde{A}_4$ asymptotes zero dual magnetic-type holomies thanks to parity. A rerun of the arguments
in~\ref{LSECTION} shows that while $\left<L({\bf x})\right>=0$ in (\ref{L00})
as expected in an Euclidean and confining thermal state, 
its dual does not vanish, i.e.

\be
\left<{\tilde L}({\bf x})\equiv {\rm Tr}\left(e^{i\frac{4\pi}{gN_c}\int_0^\beta Q\tilde{A}^Q_4(x) d\tau}\right)\right>=1
\label{HH1}
\ee
again modulo ${\cal O}(\alpha_s)$ Coulomb corrections. This behavior is consistent with the one reported on the lattice
for $N_c=2,3$~\cite{DIGIA}.

\section{Conclusions}

The central theme in this paper is non-perturbative gauge theory for temperatures in the range $(0.5-1)\,T_c$
modeled by a dense plasma of instanton-dyons. The new element in our discussion
is the introduction of the leading classical  ${\cal O}(1/\alpha_s)$ 
 interactions between the dyons and anti-dyons as recently obtained in~\cite{LARSEN-SHURYAK}
 using the classical ``streamline" set of configurations
 for $M\bar{M},L\bar{ L}$ pairs. We have assumed that
 the $M\bar{ L}, L\bar{ M}$ channels are repulsive and opposite in sign to 
 the streamline interaction. While carrying this work, this assumption has now been confirmed 
 numerically~\cite{PRIVATE}.  Another important element of our analysis is the one-loop measure of the 
 dyon and anti-dyon moduli space, in the form proposed by Diakonov and Petrov \cite{DP}.
 It leads to a small moduli space volume and thus repulsive interaction at higher density,
 which however can be made much less repulsive by introducing correlaltions between the charges.
 
 On general grounds, an ensemble of  instanton-dyons is a strongly coupled
 plasma, with significant correlations between the particles. Therefore, the statistical mechanics
 of a generic instanton-dyon ensemble  is very nontrivial and remains unsolved.
 However -- and this is the main argument of the paper -- when the plasma is dense enough
 for temperatures below $T_c$, it generates a
 large  screening mass $M$ which screens the interaction. A standard weak coupling
 plasma theory, in a form similar to the Debye-Huckel theory is then applicable . The dimensionless 
 3-density of each dyon species $n_D/4$
 in the regime considered is in the range of $n_D/4\approx T^3/4$, in agreement with the
 qualitative arguments in~\cite{SHURYAK}.
  
  Using it, we get a number of results concerning the details of the non-perturbative gauge fields,
  in the temperature range $(0.5-1)\,T_c$. First ,  in the presence of strong screening the minimum of the
  free energy  is still at the confining  (center symmetric) value of $\nu=1/2$,  with 
a vanishing Polyakov line $\left<L\right>\approx  {\rm cos}(2\pi\nu)=0$ . Second, a re-summation of the the
 linearized screening effects yields Debye-Huckel type corrections to
  the pressure and dyonic densities. We have also analyzed the  topological susceptibility,
  the gluonic compressibility, and the electric and magnetic gluonic condensates in this linearized approximation.
  
  We have calculated also the electric and magnetic screening masses, generated by the dyon ensemble.
  We have found that the latter are larger than the former in the confined phase. This is qualitatively consistent with the existing lattice data, 
  which however are much better measured for the SU(3) gauge theory rather than  the SU(2) one we have studied here.
  Finally, we have calculated  the structure factors
  in the electric and magnetic sector in the linearized screening approximation as well.
  For an estimate of the transition  temperature from $\nu =1/2$ (confinement) to $\nu =0$ (deconfinement)
  we have switched the perturbative (GPYW)  holonomy potential~\cite{WEISS}
  in section~\ref{sec_deconf}. For SU(2) the transition is observed to take place at $T_c/\sqrt{\sigma_E}\approx 0.88$.
  
  In the dyonic plasma the large spatial Wilson loops exhibit area law, while the spatial t$^\prime$ Hooft loops
  are found to be 1 modulo ${\cal O}(\alpha_s)$ Coulomb-like self-energy corrections. These dual behaviors
  were argued in~\cite{HLOOP} for confining gauge theories at zero temperature. We found them to hold
  in the confining dyon-ensemble in the regime $0.5<T<T_c$.

  Needless to say, that all these  predictions can and should be confronted with the lattice data
  in the corresponding temperature range.
     


  Finally, let us speculate about the dyon ensemble beyond the validity domain of the  Debye-Huckel
approximation. First of all, strongly coupled Coulomb plasmas are tractable by certain  analytic and/or
numerical (molecular dynamics) methods, see Refs~\cite{DH,CHO} for similar development.
Another option is to use brut force numerical simulations of the dyon ensemble \cite{LS_stat}.
Qualitatively, sufficiently strongly coupled plasmas develop either (i) correlations between particles,
resembling either  a
 liquid with crystal-like correlations (``molten salt"), or (ii) particular neutral clusters, the simplest of which
 can be the $LM$ instantons themselves or $LM\bar{L}\bar{M}$ ``instanton molecules". Recent (unquenched) lattice simulations indicate that the  instantons and
anti-instantons recombine into topologically neutral molecules across the transition temperature
~\cite{SHARMA,MOLECULE}. At much higher temperature,
the perturbative gluons dwarf all classical gauge configurations forcing the holonomy to zero.
 
 One obvious extension of this work should be into the  large number of colors
$N_c$. Strong correlations can appear, since $\Gamma_{D\bar D}\approx 1/\alpha_s\approx  N_c\gg 1$. 
Similar mechanism, leading to crystallization
appears to take place in dense holographic matter where the baryons as instantons
in the holographic direction split  into a pair of dyons and re-arrange in salt crystals~\cite{SIN}.

Another obvious extension of this work is  to include fermions, which we turn to in the second paper of the series~\cite{SECOND}.

\section{Acknowledgements}

We would like to thank Jeff Greensite, Tin Suleijmanpasic, Mithat Unsal  and Ariel Zhitnitsky
for their comments on the manuscript after it was posted. 
This work was supported by the U.S. Department of Energy under Contracts No.
DE-FG-88ER40388.

 \vfil

\begin{thebibliography}{99} \frenchspacing






\bibitem{ALL}
  T.~Schafer and E.~V.~Shuryak,
  Rev.\ Mod.\ Phys.\  {\bf 70}, 323 (1998)
  [hep-ph/9610451];
  D.~Diakonov,
  Prog.\ Part.\ Nucl.\ Phys.\  {\bf 51}, 173 (2003)
  [hep-ph/0212026];
  M.~A.~Nowak, M.~Rho and I.~Zahed,
  Singapore, Singapore: World Scientific (1996) 528 p










\bibitem{SPIN}
  N.~I.~Kochelev,
  Phys.\ Lett.\ B {\bf 426}, 149 (1998)
  [hep-ph/9610551];
  D.~Ostrovsky and E.~Shuryak,
  Phys.\ Rev.\ D {\bf 71}, 014037 (2005)
  [hep-ph/0409253];
  Y.~Qian and I.~Zahed,
  Phys.\ Rev.\ D {\bf 86}, 014033 (2012)
  [Erratum-ibid.\ D {\bf 86}, 059902 (2012)]
  [arXiv:1112.4552 [hep-ph]];
  Y.~Qian and I.~Zahed,
  Phys.\ Rev.\ D {\bf 90}, no. 11, 114012 (2014)
  [arXiv:1404.6270 [hep-ph]].








\bibitem{KVLL}
Kraan-Van-Baal NPB 533 1998
  T.~C.~Kraan and P.~van Baal,
  Nucl.\ Phys.\ B {\bf 533}, 627 (1998)
  [hep-th/9805168];
  T.~C.~Kraan and P.~van Baal,
  Phys.\ Lett.\ B {\bf 435}, 389 (1998)
  [hep-th/9806034];
  K.~M.~Lee and C.~h.~Lu,
  Phys.\ Rev.\ D {\bf 58}, 025011 (1998)
  [hep-th/9802108].








\bibitem{DP}
  D.~Diakonov and V.~Petrov,
  Phys.\ Rev.\ D {\bf 76}, 056001 (2007)
  [arXiv:0704.3181 [hep-th]];
  D.~Diakonov and V.~Petrov,
  Phys.\ Rev.\ D {\bf 76}, 056001 (2007)
  [arXiv:0704.3181 [hep-th]].
  D.~Diakonov and V.~Petrov,
  AIP Conf.\ Proc.\  {\bf 1343}, 69 (2011)
  [arXiv:1011.5636 [hep-th]];
  D.~Diakonov,
  arXiv:1012.2296 [hep-ph].

\bibitem{DPX}
  D.~Diakonov, N.~Gromov, V.~Petrov and S.~Slizovskiy,
  Phys.\ Rev.\ D {\bf 70}, 036003 (2004)
  [hep-th/0404042].


\bibitem{ZHITNITSKY}
  A.~R.~Zhitnitsky,
  hep-ph/0601057;
  S.~Jaimungal and A.~R.~Zhitnitsky,
  hep-ph/9905540;
  A.~Parnachev and A.~R.~Zhitnitsky,
  Phys.\ Rev.\ D {\bf 78} (2008) 125002
  [arXiv:0806.1736 [hep-ph]];
  A.~R.~Zhitnitsky,
  Nucl.\ Phys.\ A {\bf 921} (2014) 1
  [arXiv:1308.0020 [hep-ph]].


\bibitem{FATEEV}
  V.~A.~Fateev, I.~V.~Frolov and A.~S.~Shvarts,
  Nucl.\ Phys.\ B {\bf 154}, 1 (1979);
  B.~Berg and M.~Luscher,
  Commun.\ Math.\ Phys.\  {\bf 69}, 57 (1979).



\bibitem{SIMONOV}
  B.~Martemyanov, S.~Molodtsov, Y.~Simonov and A.~Veselov,
  JETP Lett.\  {\bf 62}, 695 (1995)
  [Pisma Zh.\ Eksp.\ Teor.\ Fiz.\  {\bf 62}, 679 (1995)];
  B.~V.~Martemyanov, S.~V.~Molodtsov, Y.~A.~Simonov and A.~I.~Veselov,
  Phys.\ Atom.\ Nucl.\  {\bf 60}, 490 (1997)
  [Yad.\ Fiz.\  {\bf 60}, 565 (1997)];
  Y.~A.~Simonov,
  In *Varenna 1995, Selected topics in nonperturbative QCD* 339-364
  [hep-ph/9509403];
  A.~Gonzalez-Arroyo and Y.~A.~Simonov,
  Nucl.\ Phys.\ B {\bf 460}, 429 (1996)
  [hep-th/9506032].

\bibitem{THOOFT}
  G.~'t Hooft,
  Nucl.\ Phys.\ B {\bf 138}, 1 (1978).






\bibitem{MAND}
  S.~Mandelstam,
  Phys.\ Rev.\ D {\bf 19}, 2391 (1979).




\bibitem{SEIBERG}
  N.~Seiberg and E.~Witten,
  Nucl.\ Phys.\ B {\bf 426}, 19 (1994)
  [Erratum-ibid.\ B {\bf 430}, 485 (1994)]
  [hep-th/9407087].

\bibitem{POLYAKOV}
  A.~M.~Polyakov,
  Phys.\ Lett.\ B {\bf 59}, 82 (1975);
  A.~M.~Polyakov,
  Nucl.\ Phys.\ B {\bf 120}, 429 (1977).



\bibitem{UNSAL1} 
  M.~Unsal and L.~G.~Yaffe,
  Phys.\ Rev.\ D {\bf 78}, 065035 (2008)
  [arXiv:0803.0344 [hep-th]];
  M.~Unsal,
  Phys.\ Rev.\ D {\bf 80}, 065001 (2009)
  [arXiv:0709.3269 [hep-th]].

\bibitem{HOLGER}
  T.~H.~Hansson, H.~B.~Nielsen and I.~Zahed,
  Nucl.\ Phys.\ B {\bf 451}, 162 (1995)
  [hep-ph/9405324].

\bibitem{UNSALALL}
  E.~Poppitz, T.~Schäfer and M.~Unsal,
  JHEP {\bf 1210}, 115 (2012)
  [arXiv:1205.0290 [hep-th]];
  E.~Poppitz and M.~Unsal,
  JHEP {\bf 1107} (2011) 082
  [arXiv:1105.3969 [hep-th]].


\bibitem{UNSAL}
  E.~Poppitz, T.~Sch�fer and M.~�nsal,
  JHEP {\bf 1303}, 087 (2013)
  [arXiv:1212.1238].


















\bibitem{WEISS}
  D.~J.~Gross, R.~D.~Pisarski and L.~G.~Yaffe,
  Rev.\ Mod.\ Phys.\  {\bf 53}, 43 (1981).
  N.~Weiss,
  Phys.\ Rev.\ D {\bf 25}, 2667 (1982);




\bibitem{TIN}
  E.~Shuryak and T.~Sulejmanpasic,
  Phys.\ Lett.\ B {\bf 726} (2013) 257
  [arXiv:1305.0796 [hep-ph]].

\bibitem{SHURYAK}
  E.~Shuryak and T.~Sulejmanpasic,
  Phys.\ Rev.\ D {\bf 86}, 036001 (2012)
  [arXiv:1201.5624 [hep-ph]];

\bibitem{FACCIOLI}
  P.~Faccioli and E.~Shuryak,
  Phys.\ Rev.\ D {\bf 87}, no. 7, 074009 (2013)
  [arXiv:1301.2523 [hep-ph]].










\bibitem{LATTICE}
  F.~Bruckmann, S.~Dinter, E.~M.~Ilgenfritz, M.~Muller-Preussker and M.~Wagner,
  Phys.\ Rev.\ D {\bf 79}, 116007 (2009)
  [arXiv:0903.3075 [hep-ph]];
  F.~Bruckmann, S.~Dinter, E.~M.~Ilgenfritz, B.~Maier, M.~Muller-Preussker and M.~Wagner,
  Phys.\ Rev.\ D {\bf 85}, 034502 (2012)
  [arXiv:1111.3158 [hep-ph]].

\bibitem{LARSEN-SHURYAK}
  R.N.~Larsen and E.~Shuryak,
  arXiv:1408.6563 [hep-ph].

\bibitem{DH}
  B.~A.~Gelman, E.~V.~Shuryak and I.~Zahed,
  Phys.\ Rev.\ C {\bf 74}, 044909 (2006)
  [nucl-th/0605046];
  S.~Cho and I.~Zahed,
  Phys.\ Rev.\ C {\bf 79}, 044911 (2009)
  [arXiv:0812.1736 [nucl-th]];
  S.~Cho and I.~Zahed,
  Phys.\ Rev.\ C {\bf 80} (2009) 014906
  [arXiv:0812.1741 [nucl-th]].



\bibitem{CHO}
  S.~Cho and I.~Zahed,
  Phys.\ Rev.\ C {\bf 82}, 054907 (2010)
  [arXiv:0910.2666 [nucl-th]];
  S.~Cho and I.~Zahed,
  Phys.\ Rev.\ C {\bf 82}, 044905 (2010)
  [arXiv:0909.4725 [nucl-th]];
  V.~S.~Filinov, Y.~B.~Ivanov, V.~E.~Fortov, M.~Bonitz and P.~R.~Levashov,
  Phys.\ Rev.\ C {\bf 87}, no. 3, 035207 (2013)
  [arXiv:1210.2664 [nucl-th]];
  V.~S.~Filinov, Y.~B.~Ivanov, M.~Bonitz, V.~E.~Fortov and P.~R.~Levashov,
  Phys.\ Lett.\ A {\bf 376}, 1096 (2012)
  [arXiv:1203.2191 [hep-ph]];
  V.~S.~Filinov, Y.~B.~Ivanov, M.~Bonitz, P.~R.~Levashov and V.~E.~Fortov,
  Phys.\ Atom.\ Nucl.\  {\bf 74}, 1364 (2011)
  [arXiv:1006.3390 [nucl-th]].




\bibitem{ADAMI}
  C.~Adami, T.~Hatsuda and I.~Zahed,
  Phys.\ Rev.\ D {\bf 43}, 921 (1991).


\bibitem{FISHER}
M.~Fisher and Y.~Levin, 
Phys.\ Rev. \ Lett.\  {\bf 71}, 3826 (1993) and references therein.




\bibitem{BORN}
  V.~G.~Bornyakov and V.~K.~Mitrjushkin,
  Phys.\ Rev.\ D {\bf 84}, 094503 (2011)
  [arXiv:1011.4790 [hep-lat]].

\bibitem{ALPHA}
  O.~Kaczmarek, F.~Karsch, F.~Zantow and P.~Petreczky,
  Phys.\ Rev.\ D {\bf 70}, 074505 (2004)
  [Erratum-ibid.\ D {\bf 72}, 059903 (2005)]
  [hep-lat/0406036].


\bibitem{TEPER}
  B.~Lucini, M.~Teper and U.~Wenger,
  JHEP {\bf 0502}, 033 (2005)
  [hep-lat/0502003].

\bibitem{TODA}
  T.~J.~Hollowood,
  hep-th/9110010.


\bibitem{GREEN}
  J.~Greensite and R.~H�llwieser,
  arXiv:1411.5091 [hep-lat].

\bibitem{GREEN1}
  J.~Greensite, B.~Lucini and A.~Patella,
  Phys.\ Rev.\ D {\bf 83}, 125019 (2011)
  [arXiv:1101.5344 [hep-th]].

\bibitem{HLOOP}
  G.~'t Hooft,
  Nucl.\ Phys.\ B {\bf 138}, 1 (1978).


\bibitem{DIGIA}
  L.~Del Debbio, A.~Di Giacomo and B.~Lucini,
  Nucl.\ Phys.\ B {\bf 594}, 287 (2001)
  [hep-lat/0006028];
  L.~Del Debbio, A.~Di Giacomo and B.~Lucini,
  Phys.\ Lett.\ B {\bf 500}, 326 (2001)
  [hep-lat/0011048].
  
  
  
  

 \bibitem{PRIVATE}
 R.~N.~Larsen, private communication.


  \bibitem{LS_stat}
   R~.N.~Larsen and E.~Shuryak, �Interacting Ensemble of Instanton-dyons 
and Confinement in SU(2) Gauge Theory�, in preparation.


\bibitem{KOSTAS}
  C.~Korthals-Altes, A.~Kovner and M.~A.~Stephanov,
  Phys.\ Lett.\ B {\bf 469}, 205 (1999)
  [hep-ph/9909516].

\bibitem{HUGO}
  H.~Reinhardt,
  Phys.\ Lett.\ B {\bf 557}, 317 (2003)
  [hep-th/0212264].

\bibitem{SHARMA}
  S.~Sharma, V.~Dick, F.~Karsch, E.~Laermann and S.~Mukherjee,
  arXiv:1311.3943 [hep-lat].

\bibitem{MOLECULE}
  E.~V.~Shuryak,
  Phys.\ Lett.\ B {\bf 196}, 373 (1987).
  T.~Schaefer, E.~V.~Shuryak and J.~J.~M.~Verbaarschot,
  Phys.\ Rev.\ D {\bf 51} (1995) 1267
  [hep-ph/9406210].
  
   \bibitem{SIN}
  M.~Rho, S.~J.~Sin and I.~Zahed,
  Phys.\ Lett.\ B {\bf 689}, 23 (2010)
  [arXiv:0910.3774 [hep-th]];
  V.~Kaplunovsky and J.~Sonnenschein,
  JHEP {\bf 1404} (2014) 022
  [arXiv:1304.7540 [hep-th]];
  S.~Bolognesi and P.~Sutcliffe,
  J.\ Phys.\ A {\bf 47}, 135401 (2014)
  [arXiv:1311.2685 [hep-th]].


\bibitem{SECOND}
Y.~Liu, E.~Shuryak and I.~Zahed, Light quarks in the screened Dyon-Anti-Dyon
Coulomb Liquid Model II, arXiv:1503.09148.










\end{thebibliography}
\end{document}